\documentclass[aps,prb,twocolumn,groupedaddress,floats,showpacs]{revtex4}

\usepackage{amsmath}
\usepackage[dvips]{graphics}
\usepackage[hypertex]{hyperref}

\newcommand{\e}{\varepsilon}
\newcommand{\w}{\omega}

\newcommand{\Ima}{\mbox{Im}}
\newcommand{\Rea}{\mbox{Re}}

\newcommand{\ff}{f}

\newcommand{\al}{\alpha}
\newcommand{\be}{\beta}
\newcommand{\ab}{\alpha \beta}

\newcommand{\gc}{g_{12}}
\newcommand{\go}{g_{11}'}
\newcommand{\gt}{g_{22}'}
\newcommand{\gts}{g_{12}'}
\newcommand{\gho}{g_{{\rm H},1}}
\newcommand{\ght}{g_{{\rm H},2}}

\newcommand{\fo}{f_{11}'}
\newcommand{\ft}{f_{22}'}
\newcommand{\fc}{f_{12}}

\newcommand{\Gcl}{G_{\rm cl}}
\newcommand{\gcl}{g_{\rm cl}}

\newcommand{\ND}{\mathcal{N}_{\rm D}} 

\newcommand{\Mi}{M_{i}}
\newcommand{\Mj}{M_{j}}
\newcommand{\Di}{\Delta_{i}}
\newcommand{\Dj}{\Delta_{j}}

\newcommand{\tD}{D}
\newcommand{\tC}{C}

\newcommand{\vF}{v}

\newcommand{\ps}{\hat{\psi}^{\phantom{\dagger}}}

\newcommand{\psd}{\hat{\psi}^{\dagger}}

\newcommand{\tr}{\mbox{ tr}}

\newcommand{\tp}{\tau_+}
\newcommand{\tm}{\tau_-}
\newcommand{\tauo}{\tau_1}
\newcommand{\taut}{\tau_2}

\newcommand{\GR}{G^{\rm R}}
\newcommand{\GA}{G^{\rm A}}
\newcommand{\LR}{L^{\rm R}}
\newcommand{\LA}{L^{\rm A}}

\begin{document}

\title{Temperature and magnetic-field dependence of the quantum
  corrections to the conductance of a network of quantum dots}
\author{Joern N.\ Kupferschmidt and Piet W.\ Brouwer}
\affiliation{Laboratory of Atomic and Solid State Physics, Cornell
  University, Ithaca, NY 14853-2501, USA}
\altaffiliation{present address}
\affiliation{Arnold Sommerfeld Center for Theoretical Physics, Ludwig-Maximilians-Universit\"at, 80333 M\"unchen, Germany}
\date{\today}

\begin{abstract}
We calculate the magnetic-field and temperature dependence of all
quantum corrections to the ensemble-averaged conductance of a network
of quantum dots. We consider the limit that the dimensionless
conductance of the network is large, so that the quantum corrections
are small in comparison to the leading, classical contribution to the
conductance. For a quantum dot network the conductance and its 
quantum corrections can 
be expressed solely in terms of the conductances and form factors 
of the contacts and the capacitances of the quantum dots. 
In particular, we calculate the 
temperature dependence of the
weak localization correction and show that it is described by an 
effective
dephasing rate proportional to temperature.
\end{abstract}

\pacs{73.23.-b, 05.45.Mt, 73.20.Fz}

\maketitle

\section{Introduction}

The low temperature conductivity of disordered metals or
 semiconductors is dominated
by the elastic scattering of electrons off impurities and
defects. While the conductivity is determined by Drude-Boltzmann
theory for not too low temperatures, quantum corrections to the
conductivity become important at temperatures low enough that the
electronic phase remains well defined over distances large in
comparison to the elastic mean free
 path.\cite{altshuler1985a,altshuler1995,imry2002} 
One usually distinguishes
two quantum corrections, the weak localization correction and the
interaction correction.\cite{anderson1979,gorkov1979,altshuler1979c} 
The former is caused by the constructive 
interference of electrons traveling along time-reversed 
paths, whereas the interaction
correction can be understood in terms of resonant scattering off
Friedel oscillations near impurities.\cite{aleiner1999,zala2001}

Although they are small in comparison to the Drude conductivity, the
quantum corrections are important because they strongly depend on
temperature and an applied magnetic field, whereas the Drude
conductivity does not (as long as impurity scattering is the dominant
source of scattering).
Theoretically, the temperature and magnetic-field dependences of the 
corrections can be expressed in terms of
the sample's diffusion constant (or, equivalently, the elastic mean free
path), which can be obtained independently from a measurement of the
Drude conductivity. The availability of quantitative theoretical
predictions has made a detailed comparison between theory and
experiment possible.\cite{bouchiat1995,pierre2003,endnote83}

The same quantum corrections also exist for a `quantum dot', a 
conductor coupled to electron reservoirs via artificial constrictions
({\em e.g.,} tunnel barriers or point contacts), such
that the conductance of the device is dominated by the contacts and
not by scattering off impurities or defects inside the sample. The
latter condition is satisfied if the product $E_{\rm Th} \nu$ of the
dot's `Thouless energy' and its density of states is much larger than
the dimensionless conductance of the contacts connecting the dot to
source and drain reservoirs. (The Thouless energy
  is the inverse of the time needed for ergodic exploration of the
  quantum dot.)

In this article we consider `open' quantum
dots, which have contact conductances larger than the conductance
quantum $e^2/h$. Because transport through a quantum dot is
dominated by the contacts, it is described by the sample's
conductance, not its conductivity. The quantum corrections then
pertain to the conductance after averaging
over an ensemble of quantum dots that differ,
{\em e.g.}, in their shape or precise impurity configuration.

While the magnetic-field dependence of quantum corrections to the
ensemble averaged conductance is in apparent agreement with the
theory,\cite{marcus1997}
the situation regarding the temperature dependence is more
complicated and no good
agreement has been reported to date. Theoretically, the temperature 
dependence of the weak localization correction to the conductance of a 
quantum dot is described by means of a `dephasing rate'
$\gamma_{\phi}$. For a quantum dot, one expects 
\begin{equation}
  \gamma_{\phi} = c T^2/E_{\rm Th}^2 \nu,
  \label{eq:sivan}
\end{equation}
where $T$ is the temperature and $c$ is a
numerical constant that depends on the dot's size and 
shape.\cite{sivan1994,blanter1996,endnote84}
The proportionality
constant $c$ can not be measured independently, however, which
is an important difference with the case of a diffusive conductor. The
absence of a separate method to determine this constant poses a 
significant difficulty when comparing theory and
experiment. A second difficulty is
the lack of a direct theory of the temperature dependence of weak
localization. Instead, the available theoretical descriptions employ a
phenomenological description
\cite{marcus1993,clarke1995,baranger1995,brouwer1995a,beenakker2005,forster2007}
and match the dephasing rate to Eq.\ (\ref{eq:sivan}), from which the
temperature dependence of weak localization can be obtained.

\begin{figure}
\includegraphics{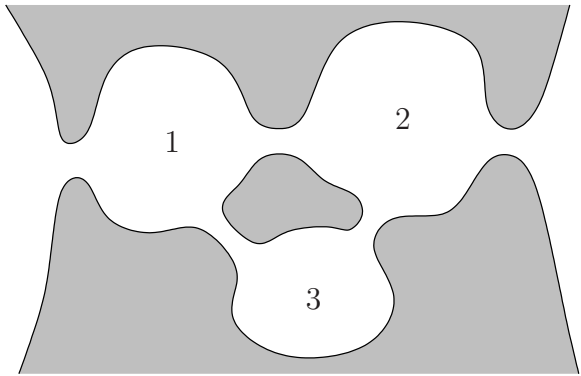}
  \caption{An example of a quantum dot network with $\ND = 3$ quantum
    dots. The conductance of the network is dominated by the
    conductances of the contacts between the dots. We assume that all
    dots in the network are `open', {\em i.e.}, all
    contact conductances are much larger than the conductance quantum
    $ e^2/h$.\label{fg:network}}
\end{figure}

In this article, we study the quantum corrections to the conductance
in a network of quantum dots or ``quantum circuit''.\cite{nazarov1994} 
(See Fig.\ \ref{fg:network} for an
example of a quantum dot network with $\ND = 3$ dots.)
Replacing a single quantum dot by a
network solves both difficulties mentioned above: A quantum
dot network allows a calculation of the complete temperature
dependence of the quantum corrections to the conductance without the
need of an intermediate step involving a phenomenological dephasing
rate and without parameters that can not be measured
independently. 
The relevant parameters in a quantum dot network
are the conductances and form factors of the contacts in the network and
the capacitances of the quantum dots.\cite{endnote85}

Our main result is an expression for the ensemble average of the
dimensionless conductance 
\begin{equation}
  G = \frac{d_{\rm s}  e^2}{h} g, 
\end{equation}
where $ d_{\rm s} = $ 1 or 2 in the absence or presence of spin 
degeneracy, respectively.
The result becomes exact in
the limit that the contact conductances are much larger than the
conductance quantum $e^2/h$,
\begin{equation}
  \langle g \rangle =
  g^{\rm cl} + \delta g^{\rm WL} + \delta g^{{\rm int},1} +
  \delta g^{{\rm int},2}.
  \label{eq:gsepavg}
\end{equation}
Here $g^{\rm cl}$ is the `classical' conductance one obtains from
Drude-Boltzmann theory, while $\delta g^{\rm WL}$, $\delta g^{{\rm
    int},1}$, and $\delta g^{{\rm int},2}$ are three quantum
corrections to $\langle g \rangle$. Explicit expressions for $g^{\rm
  cl}$ and the three quantum corrections in terms of the contact
conductances and the capacitances of the quantum dots in the network,
as well as the precise
conditions for the validity of Eq.\ (\ref{eq:gsepavg}) will be given 
in Sec.\ \ref{sec:network} below. The correction
$\delta g^{\rm WL}$ is the weak localization correction. It is the
only quantum correction that is affected by the application of a
magnetic field. The remaining two corrections arise from
electron-electron interactions. 
The first interaction correction $\delta
g^{{\rm int},1}$ represents a non-local correction to the
conductance that exists for networks of two or more quantum dots
only.\cite{beloborodov2003b,golubev2004b,brouwer2008a} It is the
counterpart of the Altshuler-Aronov correction in the theory of
disordered conductors. The second correction,
 $\delta g^{{\rm int},2}$, describes 
the renormalization of the contact conductances
by the interactions. It is usually referred to as (dynamical) Coulomb 
blockade, an effect that is well-known from the theory of
transport through tunnel junctions in series with a
high impedance or quantum dots with tunneling 
contacts.\cite{devoret1990,girvin1990,nazarov1989b,flensberg1993,furusaki1995a,furusaki1995b,golubev2001,levyyeyati2001,kindermann2003,golubev2004a,bagrets2005,brouwer2005,brouwer2005d}
Its counterpart in the theory of disordered conductors is the 
Altshuler-Aronov correction to the tunneling
density of states.\cite{altshuler1979a}

The fact that the temperature dependence of quantum corrections in a
quantum dot network does not depend on details of individual dots has
its origin in the different form of the relevant electron-electron 
interaction modes in a quantum dot network and in a single dot.
In a single quantum dot, the dominant contribution to the
electron-electron interaction is the uniform mode, the strength of
which is set
by the dot's capacitance.
Apart from a possible renormalization of the contact
conductances, $\delta g^{{\rm int},2}$,
the uniform mode has no effect on the
quantum correction to the dot's 
conductance.\cite{golubev2004a,brouwer2005,brouwer2005d,ahmadian2005} 
In particular, the weak localization 
correction $\delta g^{\rm WL}$ is unaffected by the interaction 
and the non-local interaction correction $\delta g^{{\rm int},1} $ vanishes. 
Instead, electron-electron interactions 
determine $\delta g^{\rm WL} $ and $\delta g^{{\rm int},1} $ in 
a single quantum dot through sub-dominant
non-uniform interaction modes, which are known to 
depend on
the precise sample details.\cite{sivan1994,aleiner2002}
For a quantum dot network, on the other
hand, there exist interaction modes that are uniform inside
each dot but not across the full network. With such interaction modes,
 all three interaction corrections $\delta g^{\rm WL}$, $\delta g^{\rm
 int,1}$, and $\delta g^{\rm int,2}$ are generically nonzero and
 temperature dependent.
Moreover, because these modes are
uniform inside each quantum dot, their properties depend on the
contacts between the dots and on the dot capacitances only, not on 
the precise geometry of each
dot separately. It is this essential feature that makes a quantum
dot network an ideal paradigm for studying the effect of
electron-electron interactions on quantum transport in finite-size
systems.

Separate aspects of the problem we address here have been
considered before. Weak localization in single quantum dots without
interactions has been studied by various
authors,\cite{baranger1994,jalabert1994,efetov1995,argaman1995,argaman1996,aleiner1996,brouwer1996a,vavilov1999,richter2002,whitney2007}
as well as the effect of the uniform interaction mode on the
conductances of the contacts connecting the dot to the electron
reservoirs.\cite{nazarov1989b,girvin1990,devoret1990,flensberg1993,furusaki1995a,furusaki1995b,golubev2001,levyyeyati2001,kindermann2003,golubev2004a,bagrets2005,brouwer2005,brouwer2005d}
(See Ref.\ \onlinecite{ahmadian2005} for a discussion of a
comparable effect involving spin-dependent interactions in the quantum dot.)
Also, while it is known that the uniform interaction mode has no effect on
weak localization because a spatially uniform fluctuating potential affects
phases of time-reversed trajectories in the same
way,\cite{brouwer2005,brouwer2005d} the uniform
interaction mode can suppress interference contributions to other
observables if the quantum dot is part
of an interferometer.\cite{seelig2001,seelig2003,endnote86}

Weak localization in networks of quantum dots, but
without interactions, was considered by Argaman for dots connected by
ideal contacts,\cite{argaman1995,argaman1996} and by Campagnano and
Nazarov for dots connected by arbitrary contacts.\cite{campagnano2006}
Golubev and Zaikin calculated the interaction corrections 
$\delta g^{{\rm int},1}$ and $\delta g^{{\rm int},2}$ for a linear 
array of quantum dots,\cite{golubev2004b} as well as the weak 
localization correction for non-interacting electrons (but with a 
phenomenological dephasing rate).\cite{golubev2006} In a recent
publication, the same authors also considered the full temperature 
dependence of weak localization in the special case of a double 
quantum dot (a network with $\ND = 2$ quantum dots) with tunneling 
contacts,\cite{golubev2007} and reported that electron-electron
interactions suppress weak localization even at zero temperature,
a conclusion that contradicts the common wisdom that there is no
dephasing from electron-electron interactions at zero
temperature.\cite{altshuler1985a,imry2002}

Weak localization and interaction corrections have also been 
considered for networks of 
diffusive metallic wires.\cite{texier2004,texier2007} 
Large arrays of quantum dots connected by tunneling contacts further
appear in the study of granular metals.\cite{beloborodov2007}
Beloborodov and coworkers considered the interaction corrections
$\delta g^{{\rm int},1}$ and $\delta g^{{\rm int},2}$ for a granular
metal,\cite{beloborodov2001,beloborodov2003a,beloborodov2003b,efetov2003b,beloborodov2004} but accounted for weak
localization and its temperature dependence only via a phenomenological 
dephasing rate and a renormalized diffusion
constant. A microscopic theory of the temperature dependence
of weak localization in granular metals was given by Blanter {\em et
 al.\ }in the high temperature limit.\cite{blanter2006} 
Our present analysis (as well as that of Ref.\
\onlinecite{golubev2004b}) is for contacts of arbitrary transparency 
and contains contributions to weak localization and to the interaction 
correction to the conductance that are absent in a network where all 
contacts are
tunneling contacts. Our results agree with the literature wherever
applicable, except for the zero-temperature limit of the weak
localization correction $\delta g^{\rm WL}$, where we find that
weak localization is unaffected by electron-electron interactions, in
contrast to Ref.\ \onlinecite{golubev2007}.

The remainder of our article is organized as follows. In Sec.\ 
\ref{sec:network} we introduce the relevant parameters needed to
describe the quantum dot network, formulate our main assumptions, and
present our main result, an expression for the ensemble-averaged
conductance and its quantum corrections. In Sec.\ \ref{sec:motivation} 
we motivate our result for the temperature dependence of the weak
localization correction using semiclassical arguments. In Sec.\
\ref{sec:model} we then turn to a fully quantum mechanical calculation
of the conductance and its quantum corrections using random matrix
theory. We specialize to the simplest network, a double quantum dot,
in Sec.\ \ref{sec:double} and discuss the origin of the difference
between our result and Ref.\ \onlinecite{golubev2007} 
for the zero-temperature limit of
weak localization. We conclude in Sec.\ \ref{sec:conc}. 

\section{Definition of the problem and main results}
\label{sec:network}

\subsection{Network of quantum dots}

We consider a network of $\ND$ quantum dots, coupled to two
electron reservoirs. A schematic drawing of a network is shown in
Fig.\ \ref{fg:network}. In this section we introduce the relevant
parameters to describe the quantum dot network and summarize our main
results.

The quantum dots are connected to each other
and to source and drain electron reservoirs via point contacts. The 
dots will be
labeled by an index $i=1,\ldots,\ND$; the reservoirs are labeled by
the index $a=1,2$. The contact between 
dots $i$ and $j$ is described by its dimensionless conductance
$g_{ij}$ (per spin direction)
and its form factor $f_{ij}$. Both $g_{ij}$ and $f_{ij}$ are
defined in terms of the transmission matrix $t_{ij}$ of the contact,
\begin{eqnarray}
  g_{ij} &=& \mbox{tr}\, t_{ij} t_{ij}^{\dagger}, \ \
  f_{ij} = \mbox{tr}\, (t_{ij} t_{ij}^{\dagger})^2.
  \label{eq:fgdef}
\end{eqnarray}
Form factors are related to Fano factors $\beta$ often 
encountered in the literature via 
$ \beta_{ij} = (g_{ij} - f_{ij})/g_{ij} $. 
The dimensionless conductances and form factors are symmetric, $g_{ij}
= g_{ji}$ and $f_{ij} = f_{ji}$, $i,j=1,\ldots,\ND$. 
Spin degeneracy will be explicitly taken into account 
via the parameter $d_{\rm s} = {1,2}$.

Similarly, the contacts between the $i$th quantum dot and reservoir
$a$, $a=1,2$, are described by a dimensionless conductance
$g_{ia}'=g_{ai}'$ and a form factor $f_{ia}' = f_{ai}'$, which are
related to the transmission matrix $t_{ia}'$ of these contacts as
\begin{eqnarray}
  g_{ia}' &=& \mbox{tr}\, t_{ia}' t_{ia}'^{\dagger}, \ \
  f_{ia}' = \mbox{tr}\, ( t_{ia}' t_{ia}'^{\dagger} )^2.
  \label{eq:fgpdef}
\end{eqnarray}
For ballistic contacts one has $f = g$; for tunneling contacts one has
$f \ll g $. Throughout we assume that all conductances are large,
\begin{equation}
  g_{ij},\ g_{ia}', \gg 1,\ \ i,j=1,\ldots,\ND,\ \ a=1,2.
  \label{eq:gineq}
\end{equation}
(One may replace this condition by the less strict requirement
that each quantum dot be well connected to one of the
 two reservoirs, such that the regime of strong Coulomb blockade is avoided.)
For future use, we arrange the conductances and form factors in 
$\ND \times \ND$ 
matrices $\tilde g$ and $\tilde f$ with elements
\begin{eqnarray}
  \tilde{g}_{ij} &  =  & \left\{ \begin{array}{ll} \sum_{a=1}^{2}
  g_{ aj}' + \sum_{k \neq i}^{\ND } g_{ ik} &  i = j,\\ -
  g_{ ij} & i \neq j \mbox{,}  \end{array} \right.  
  \label{eq:tildeg} \\
  \tilde{f}_{ij} &  =  & \left\{ \begin{array}{ll} \sum_{a=1}^{2}
  f_{ aj}' + \sum_{k \neq i}^{\ND } f_{ ik} &  i = j,\\ -
  f_{ ij} & i \neq j \mbox{.}  \end{array} \right.
  \label{eq:tildef}
\end{eqnarray}

The quantum dots are assumed to be disordered 
or ballistic-chaotic, with density of states $\nu_i$ per spin degree 
of freedom and Thouless energy $E_{{\rm Th},i}$, $i=1,\ldots,\ND$.
The Thouless energy $ E_{{\rm Th},i} = \hbar/ \tau_{{\rm erg},i} $, 
where $\tau_{{\rm erg},i} $ is the time for ergodic exploration of the 
$i$th quantum dot. If the electron motion is diffusive inside each quantum
dot with diffusion constant $D$, $E_{{\rm Th},i} \sim D/L_i^2$ where $L_i$ 
is the linear size of dot $i$.
(Our definition, while common in the literature, differs from 
some references where $E_{{\rm Th},i}$ is the inverse of the dot's dwell time.)
We assume
\begin{equation}
  E_{{\rm Th},i} \nu_i \gg \tilde g_{ii},\ \ i=1,\ldots,\ND,
  \label{eq:RMTineq}
\end{equation}
so that random matrix theory can be used to describe the electronic
states in the quantum dot network.
An external magnetic field is
described by means of the dimensionless numbers
\begin{equation}
  g_{{\rm H},i} = E_{{\rm Th},i} \nu_i \frac{\Phi_i^2}{\Phi_0^2}, \ \
  i=1,\ldots,\ND,
  \label{eq:gH}
\end{equation}
where $\Phi_i$ is the magnetic flux through the $i$th quantum dot and
$\Phi_0 = hc/e$ is the flux quantum. In order to simplify the notation,
we arrange the densities of states $\nu_i$ and the parameters 
$g_{{\rm H},i}$ in diagonal $\ND$-dimensional matrices $\tilde \nu$ and $\tilde
g_{\rm H}$,
\begin{equation}
  \tilde \nu_{ij} = \nu_{i} \delta_{ij},\ \ 
  (\tilde{g}_{\rm H})_{ij} = g_{{\rm H},i} \delta_{ij}, \ \
  i,j=1,\ldots,\ND.  
  \label{eq:tildenu}
\end{equation}
Corrections to the conductance that depend on the magnetic 
field will only be relevant where  $g_{{\rm H},i}$ is of order
$\tilde g_{ii}$ or less, otherwise they will be fully suppressed. 
In that parameter range, the flux through
the insulating regions between the quantum dots is much smaller than
$\Phi_0$, so that the corresponding Aharonov-Bohm phases can be
neglected.

The inequality (\ref{eq:RMTineq}) also implies that the
electron-electron interaction in each dot is well screened.\cite{aleiner2002}
 Hence, the electron-electron
interaction couples to the total charge $q_i = e n_i$ of each dot
only. Such an
interaction is described by means of capacitances $C_{ij}$ for the
capacitive coupling between dots (if $i \neq j$) and for each dot's 
self-capacitance (if $i = j$).
Again, we arrange the capacitances into
an $\ND$-dimensional matrix $\tilde C$, 
\begin{eqnarray}
  \tilde{C}_{ij} & = & \left\{ \begin{array}{ll} \sum_{k=1}^{\ND}
  C_{ik} &  i = j,\\ - C_{ij} & i \neq j. \end{array}  \right.  
  \label{eq:Cdef}
\end{eqnarray}
For metallic dots, one has the inequality
\begin{equation}
  \tilde C_{ii}/e^2 \ll \nu_i,\ \ i=1,\ldots,\ND.
  \label{eq:Cineq}
\end{equation}

\subsection{Quantum corrections to the conductance}

Our main result is a calculation of the ensemble-averaged
conductance $\langle G \rangle =  (d_{\rm s} e^2/h) \langle g \rangle$ of
the quantum dot network as a function of temperature,
$$
  \langle g \rangle = g^{\rm cl} + 
  \delta g^{\rm WL} + \delta g^{{\rm int},1} +
  \delta g^{{\rm int},2},
$$
where $g^{\rm cl}$ is the classical conductance of the network and
$\delta g^{\rm WL}$, $\delta g^{{\rm int},1}$, and $\delta g^{{\rm
int},2}$ are corrections. 
The average conductance is calculated using
the following limiting procedure for the parameters of the network:
\begin{enumerate}
\item We first take the limit (\ref{eq:RMTineq}) needed for the
applicability of random matrix theory, while keeping the ratios
$\nu_i/\nu_j$ and $T/\nu_i$, as well as the $g_{{\rm H},i}$ fixed, 
$i,j=1,\ldots,\ND$.
\item We then take the limit (\ref{eq:gineq}) of large contact 
conductances, while keeping the ratios $g_{ij}/g_{ik}$,
$g_{ij}/g_{{\rm H},i}$, and $g_{ij}/g_{ia}'$ fixed,
$i,j,k=1,\ldots,\ND$, $a=1,2$.
\item Finally, we simplify our results using the inequality
  (\ref{eq:Cineq}), if possible.
\end{enumerate}
In all three limiting steps, the number $\ND$ of dots in the network is
kept constant. Keeping the ratio $T/\nu_i$ fixed in the first limiting
step eliminates interaction corrections from non-uniform interaction
modes inside the quantum dots, see Eq.\ (\ref{eq:sivan}) above. 
In the second limiting step, we do not 
make any assumptions about the temperature, thus allowing for the full
range of temperature-dependent effects that can be described within
random matrix theory. We note that, while the classical conductance $g^{\rm
  cl}$ diverges in this limiting procedure, this divergence
does not affect the temperature or magnetic-field  dependence of
$\langle g \rangle$ because $g^{\rm cl}$ does not depend on
temperature or magnetic field. 
Corrections not
included in Eq.\ (\ref{eq:gsepavg}) are either small in the limit
(\ref{eq:gineq})
of large contact conductances or small in the limit (\ref{eq:RMTineq}) used to justify the use of
random matrix theory.

The leading term $g^{\rm cl}$ in Eq.\ (\ref{eq:gsepavg}) reads
\begin{eqnarray}
  g^{\rm cl} &=& 
   \sum_{i,j = 1}^{\ND} g_{1 i}' ( \tilde{g}^{-1})_{ij} g_{j 2}'
  \nonumber \\ &=&
  g_{1 \cdot}' \tilde g^{-1}_{\cdot \cdot} g_{\cdot 2}',
  \label{eq:gcl}
\end{eqnarray}
where, in the second line of Eq.\ (\ref{eq:gcl}),
we have written ``$\cdot$'' to denote indices in adjacent
factors that are summed over as in matrix multiplication. [Compare
with the first line of Eq.\ (\ref{eq:gcl}).]
This shorthand notation will be employed throughout the
text.

The correction $\delta g^{\rm WL}$ is the weak localization
correction to the ensemble-averaged conductance. 
It can be distinguished from the remaining two corrections
$\delta g^{{\rm int},1}$ and $\delta g^{{\rm int},2}$
because $\delta g^{\rm WL}$ depends on an applied magnetic field
whereas $\delta g^{\rm int}_1$ and $\delta g^{{\rm int},2}$ do not.
We find
\begin{eqnarray}
  \delta g^{\rm WL}
  &=&
  2 \sum_{i,j=1}^{\ND}
  \, \tilde{c}_{ij}\, 
  g_{1\cdot}' (\tilde g_{\cdot i}^{-1} - \tilde g_{\cdot j}^{-1}) 
  (\tilde g - \tilde f)_{ij}
  \tilde g_{j\cdot}^{-1} 
  g_{\cdot 2}'
  \nonumber \\ &&  \mbox{}
  + \sum_{i=1}^{\ND} \, \tilde{c}_{ii}\, 
  (g_{1\cdot}' \tilde g_{\cdot \cdot}^{-1} \tilde f_{\cdot i} - f_{1i}')
  \tilde g_{i\cdot}^{-1} g_{\cdot 2}'
  \nonumber \\ &&  \mbox{}
  + \sum_{i=1}^{\ND} \, \tilde{c}_{ii} \, 
  g_{1\cdot}' \tilde g_{\cdot i}^{-1}
  (\tilde f_{i\cdot} \tilde g^{-1}_{\cdot\cdot} g_{\cdot 2}' - f_{i2}')
  \nonumber \\ &&  \mbox{}
  - \sum_{i,j=1}^{\ND} \tilde f_{ij} \, \tilde{c}_{jj} \,
  g_{1\cdot}' \tilde g_{\cdot i}^{-1} \tilde g_{i \cdot}^{-1} g_{\cdot 2}'
  ,
  \label{eq:dGWL}
\end{eqnarray}
where the $\ND \times \ND$ matrix $\tilde c$ is the counterpart of the
``Cooperon'' in the theory of weak localization in disordered
conductors. For the quantum dot network, $\tilde c$ reads
\begin{equation}
  \tilde c_{ij} = \sum_{k=1}^{\ND} \frac{1}{\pi
  \hbar \nu_k} (\Gamma+\Gamma_{\rm H}+\Gamma_{\phi})^{-1}_{ik,jk} ,
  \label{eq:cijdef}
\end{equation}
where $\Gamma$, $\Gamma_{\rm H}$, and $\Gamma_{\phi}$ are rank-four
tensors,
\begin{eqnarray}
  \Gamma_{ik,jl} &=&
  \frac{1}{2 \pi \hbar \nu_i} \tilde g_{ik} \delta_{jl} +
  \frac{1}{2 \pi \hbar \nu_j} \delta_{ik} \tilde g_{jl}
  \label{eq:Gamma}
  \nonumber \\
  (\Gamma_{\rm H})_{ik,jl} &=&
  \frac{1}{2 \pi \hbar \nu_i} \tilde g_{{\rm H},ik} \delta_{jl} +
  \frac{1}{2 \pi \hbar \nu_j} \delta_{ik}
  \tilde g_{{\rm H},jl},
  \nonumber \\
  (\Gamma_{\phi})_{ik,jl} &=&
  \frac{4 \pi T}{d_{\rm s} \hbar} (\tilde g^{-1}_{ii} + \tilde g^{-1}_{jj}
  - 2 \tilde g^{-1}_{ij}) \delta_{ik} \delta_{jl}.
\end{eqnarray}
The terms $\Gamma_{\rm H}$ and $\Gamma_{\phi}$ describe the
suppression of weak localization by a magnetic field and
electron-electron interactions, respectively. 
In the limit of low temperatures $\Gamma_{\phi} = 0$ and 
Eq.\ (\ref{eq:cijdef}) simplifies to
\begin{equation}
  \tilde c_{ij} = (\tilde g
+ \tilde g_{\rm H})^{-1}_{ij}.
  \label{eq:codlimit}
\end{equation}
For high temperatures $(\Gamma_{\phi})_{ii,jj}$ diverges [other elements are 
zero because of the Kronecker deltas in Eq.\ (\ref{eq:Gamma})], except for 
the diagonal elements with $i=j$. Hence, one finds
\begin{equation}
  \tilde c_{ij} \equiv 
  \tilde c_{ij}^{\rm d}
  = (\tilde g^{\rm d} + \tilde g_{\rm H})^{-1}_{ij},
  \label{eq:cdlimit}
\end{equation}
where $\tilde g^{\rm d}_{ij}$ is the diagonal part of the matrix
$\tilde g$, $\tilde g^{\rm d}_{ij} = \tilde g_{ij} \delta_{ij}$. This
is the contribution to the weak localization correction that arises
from self-returning electron trajectories that reside inside
one quantum dot only and, hence, are unaffected by dephasing from
electron-electron interactions.\cite{blanter2006}

The first interaction correction $\delta g^{{\rm int},1}$
is
\begin{widetext}
\begin{eqnarray}
  \delta g^{{\rm int},1} &=&  \frac{ 2 \pi}{d_{\rm s}}
  \int d\w 
  \left(
  \frac{\partial}{\partial \w}
  \omega \coth \frac{\w}{2 T} \right)
  \sum_{\alpha,\beta=1}^{\ND} 
  \sum_{k,l=1}^{\ND}
  \mbox{Im}\, \Big[
   \nu_{\alpha} 
  ( 2 \pi i \w \tilde g^{-1}_{\alpha\beta} - \tilde \nu^{-1}_{\alpha\beta} )
 \nu_{\beta} 
  \nonumber \\ && \mbox{} \times
  (\tilde g - 2 \pi i \tilde \nu \w)^{-1}_{\alpha k}
  (\tilde g - 2 \pi i  \tilde \nu \w)_{kl}
  (\tilde g - 2 \pi i \tilde \nu \w)^{-1}_{\beta l}
  g_{1\cdot}(\tilde g^{-1}_{\cdot \alpha} -
  \tilde g^{-1}_{\cdot k})
  ( \tilde g^{-1}_{l \cdot}  -
   \tilde g^{-1}_{\beta \cdot}) g_{\cdot 2}' \Big].
  \label{eq:AAgen1}
\end{eqnarray}
\end{widetext}
The second interaction correction $\delta g^{{\rm int},2}$ represents
the renormalization of the conductances between the quantum dots and
between the dots and the reservoirs as a result of the
electron-electron interactions,
\begin{eqnarray}
  \delta g^{{\rm int},2} &=&
  \sum_{j=1}^{\ND} 
  \sum_{a=1}^{2} \frac{\partial g_{\rm cl}}{\partial g_{aj}'}
  \delta g_{aj}' + 
  \sum_{j < k}^{\ND} \frac{\partial g_{\rm cl}}{\partial g_{jk}}
  \delta g_{jk}.
  \label{eq:AAgen2}
\end{eqnarray}
The interaction corrections $\delta g_{ia}'$ and $\delta g_{ij}$ exist
for non-ideal contacts with $f_{ij} < g_{ij}$, $f_{ia}' < g_{ia}'$
only, $i,j=1,\ldots,\ND$, $a=1,2$,
\begin{eqnarray}
  \label{eq:AAgen2a}
  \delta g_{aj}' &=& -(g_{aj}' - \ff'_{aj})
  \int \frac{d\w}{\w}
  \left( \frac{\partial}{\partial \w} \w \coth \frac{\w}{2 T} \right)
  \mbox{Re}\, \delta \tilde z_{jj}, \nonumber \\
  \delta g_{jk} &=& -(g_{jk} - \ff_{jk})
   \int \frac{d\w}{\w}
  \left( \frac{\partial}{\partial \w} \w \coth \frac{\w}{2 T} \right)
  \nonumber \\ && \mbox{} \times
  \mbox{Re}\, (\delta \tilde z_{jj}
  + \delta \tilde z_{kk} - 2 \delta \tilde z_{jk}),  
  \label{eq:AAgen2c}
\end{eqnarray}
where $\delta \tilde z$ is the difference of the network's
dimensionless
impedance
matrices with and without interactions,
\begin{equation}
  \delta \tilde z = (d_{\rm s}  \tilde g - 2 \pi i \w \tilde C / e^2 )^{-1} -
  (d_{\rm s} \tilde g - 2 \pi i \w  d_{\rm s} \tilde \nu)^{-1}.
  \label{eq:deltaz}
\end{equation}

The interaction correction $\delta g^{{\rm int},1}$ was obtained 
previously by Golubev and Zaikin for a linear array of quantum
dots,\cite{golubev2004b} and by Beloborodov {\em et al.} in the
context of a granular metal.\cite{beloborodov2003b} It 
is the counterpart of the Altshuler-Aronov correction
in disordered metals, where it arises from the diffusive dynamics of
the electrons. Although the electron dynamics is not diffusive in a
quantum dot network, it is non-ergodic, which is sufficient for
this interaction correction to appear. (The exception is a quantum 
dot network consisting
of a single quantum dot only, for which the electron motion is
ergodic. Indeed, one verifies that $\delta g^{{\rm int},1} = 0$ if $\ND
 = 1$, in agreement with Refs.\ 
\onlinecite{golubev2004a,golubev2004b,brouwer2005,brouwer2005d}.) 
A semiclassical
calculation of $\delta g^{{\rm int},1}$ for the special case of a
double quantum dot with ballistic contacts can be found in Ref.\
\onlinecite{brouwer2008a}.

For the case of a single quantum dot, the renormalization of the
contact conductances $\delta g^{{\rm int},2}$ or ``dynamical Coulomb
blockade'' was obtained 
previously in Refs.\ \onlinecite{golubev2001,levyyeyati2001,kindermann2003,golubev2004a,bagrets2005,brouwer2005,brouwer2005d}.
The renormalization of the contact conductances in the quantum dot
network is essentially the same as in the case of a single quantum dot
or a single tunnel junction coupled to a high-impedance electrical
environment --- in both cases the change of the
contact conductance is proportional to the factor $(g-f)$ ---,
the only difference being that the impedance $z$ is replaced by the
impedance matrix $\tilde z$ in the case of the quantum dot
network.\cite{golubev2004b}
The same conclusion was reached for the
interaction correction in an array of quantum dots with
tunneling contacts in the context of transport through a granular
metal.\cite{beloborodov2001,beloborodov2003a,beloborodov2003b,efetov2003b,beloborodov2004}

Equations (\ref{eq:gsepavg})--(\ref{eq:deltaz}) provide a general solution
for the ensemble-averaged conductance and its quantum corrections in
an arbitrary quantum dot network for arbitrary temperature. These
expressions can be simplified only by specializing to a particular
quantum dot network. In Sec.\ \ref{sec:double} we analyze these
expressions for the case of a double quantum dot, a
network consisting of two quantum dots. 

Although it is not possible to proceed quantitatively without
specializing to a particular network, we can compare the sizes of
these three quantum corrections and their typical temperature 
dependences. For the limiting procedure taken
here --- see the discussion following Eq.\ (\ref{eq:gsepavg}) ---, the
relevant temperature scale for dephasing of the weak localization
correction is\cite{blanter2006}
\begin{equation}
  T_{\phi} = \hbar\max(g, g_{\rm H})/\tau_{\rm D},
  \label{eq:Tphi}
\end{equation}
where 
\begin{equation}
  \tau_{\rm D} \sim \hbar \nu/g
\end{equation}
is the typical dwell time for the network. (Here $g$ and
$g_{\rm H}$ are shorthand notations for typical values of $g_{ij}$
or $g_{{\rm H},i}$ in the network, respectively.) For the 
interaction corrections $\delta g^{{\rm int},1}$ and
$\delta g^{{\rm int},2}$, the relevant temperature scales are
$\hbar/\tau_{\rm D}$ and the inverse charge relaxation time
\begin{equation}
  \hbar/\tau_{\rm c} \sim e^2 g/\hbar C.
\end{equation}
(In a more precise analysis 
one needs to identify $\ND$ dwell times and $\ND$ charge relaxation 
times for a network consisting of $\ND$ quantum dots, see Sec.\
\ref{sec:double} for an explicit calculation for $\ND = 2$.) Since,
typically, $C/e^2 \ll \nu$, the
charge relaxation time and the dwell time satisfy the inequality 
\begin{equation} 
  \tau_{\rm c} \ll \tau_{\rm D}.
\end{equation}

With these definitions, we find the order of magnitude of the
weak localization correction $\delta g^{\rm WL}$ to be
\begin{equation}
  \delta g^{\rm WL} \sim \delta g_{\rm WL}^{\rm d} + 
  \frac{\delta g_{\rm WL}^{\rm od}}{\max(1,T/T_{\phi})},
  \label{eq:gWLT}
\end{equation}
where $\delta g_{\rm WL}^{\rm d}$ and $\delta g_{\rm WL}^{\rm od}$ are 
constants of order $\min(1,g/g_{\rm H})$.
Similarly, for interaction corrections we find
\begin{eqnarray}
  \delta g^{{\rm int},1} & \sim & \min(1,\hbar/T\tau_{\rm D}), \\
  \delta g^{{\rm int},2} & \sim & \left\{
  \begin{array}{ll}
  \ln[\max(\tau_{\rm c} T/\hbar,\tau_{\rm c}/\tau_{\rm
  D})] & \mbox{if $T \ll \hbar/\tau_{\rm c}$}, \\
  \hbar/T \tau_{\rm c} &
  \mbox{if $T \gg \hbar/\tau_{\rm c}$}, \end{array} \right.
  ~~~~
  \label{eq:dgint2asymp}
\end{eqnarray}
independent of the magnetic field.
All three quantum corrections need
to be taken into account for a complete description of the temperature
and magnetic-field dependence of the conductance of a quantum dot
network.  In particular, in order to correctly describe the
temperature dependence of $\langle g \rangle$ for $T \lesssim
\hbar/\tau_{\rm D}$, $\delta g^{{\rm int},1}$ can not be neglected
with respect to $\delta g^{{\rm int},2}$, in spite of the fact that
$\delta g^{{\rm int},2}$ is larger than $\delta g^{{\rm int},1}$ by
(at least) a large logarithmic factor $\ln(\tau_{\rm D}/\tau_{\rm
c})$.

The temperature dependence (\ref{eq:gWLT}) implies a dephasing rate
that is linear in temperature. A linear temperature dependence of the
dephasing rate was obtained previously by Blanter {\em et al.}
in the context of a granular metal,\cite{blanter2006} and by Seelig
and B\"uttiker for a single quantum dot embedded in one arm of an
interferometer.\cite{seelig2001} In both cases, the linear
temperature dependence of the dephasing rate arose because the
fluctuations of the electric potential can be considered
classical, similar to the situation encountered in one-dimensional and
two-dimensional disordered conductors.\cite{altshuler1982b} As we
discuss in the following sections, the same
mechanism is responsible for the linear temperature dependence of the
dephasing rate in the quantum dot network.

In the next section we describe a semiclassical derivation of the weak
localization correction and its temperature dependence, 
Eq.\ (\ref{eq:dGWL}) above. 
A full quantum mechanical calculation of all
three corrections to the conductance is given in Sec.\
\ref{sec:model}. We apply the general results presented here
to the specific case of a double quantum dot in Sec.\
\ref{sec:double}.

\section{Weak localization: semiclassical considerations}

\label{sec:motivation}

In this section, we give a semiclassical 
argument for the temperature dependence of the weak localization
correction to the conductance of a quantum dot network. These
arguments provide a semiclassical interpretation of the fully
quantum mechanical calculations of the next section.

\begin{figure}
  \includegraphics{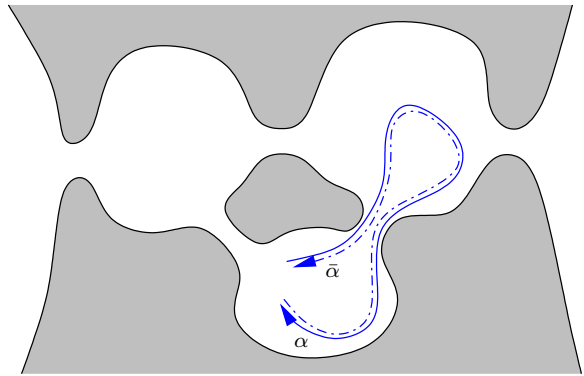}
  \caption{Schematic drawing of a trajectory $\alpha$ and its
  time-reversed $\bar \alpha$ that contribute to 
  the Cooperon propagator $\tilde c$.
  \label{fg:dd_orbit}}
\end{figure}

Weak localization appears because of constructive interference of
time-reversed trajectories. This interference leads to a small
increase to the probability $P_{\rm ret}$ that an electron returns 
to its point of origin. Following the standard
arguments,\cite{imry2002,altshuler1995} $P_{\rm ret}$ is 
calculated as a square of the return amplitude which, 
in turn, is written as a sum of amplitudes $A_{\alpha}$ over all 
returning paths $\alpha$. (These paths are classical paths in ballistic
conductors,\cite{aleiner1996,richter2002}
and quantum diffractive paths in conductors with
impurity scattering.) The quantum correction to $P_{\rm ret}$
then follows from interference between a path $\alpha$ and its
time-reversed $\bar{\alpha}$. 
Since the length of the self-returning
path is arbitrary, the weak localization correction to the
dc conductance is proportional to the time
integral of the interference correction to the return probability, 
known as the ``Cooperon'' in the diagrammatic theory of weak 
localization.\cite{imry2002,altshuler1995}
The counterpart of the Cooperon for the quantum dot
network is the quantity
\begin{equation}
  \tilde c_{ij} \sim \frac{1}{( 2 \pi \hbar)^2 \nu_i \nu_j}
  \sum_{\alpha} A_{\alpha} (A_{\bar \alpha})^*,
  \label{eq:cij0}
\end{equation}
where the sum is over all trajectories $\alpha$ that originate in dot
$j$ and end in dot $i$ and $\bar \alpha$ is the time-reversed of
$\alpha$, see Fig.\ \ref{fg:dd_orbit}. [Note that the return
probability involves the diagonal elements $\tilde c_{ii}$ of the Cooperon
matrix only. We have included non-diagonal elements in Eq.\
(\ref{eq:cij0}) above in view of the discussion of interaction effects
below. Non-diagonal elements $\tilde c_{ij}$ with $i$ and $j$ in adjacent
dots also appear for the description of weak localization in a network
of quantum dots with tunneling contacts, see Eq.\ (\ref{eq:dGWL})
above.]

At zero temperature and without a magnetic field, $A_{\bar \alpha} =
A_{\alpha}$. We may then calculate $\tilde c_{ij}$ using that
$|A_{\alpha}|^2$ is the probability that an electron propagates along
trajectory $\alpha$. Hence
\begin{eqnarray}
  \tilde c_{ij} &=&
  \frac{1}{2 \pi \hbar \nu_i}
  \int_0^{\infty} d\tau P_{ij}(\tau),
  \label{eq:cij0semi}
\end{eqnarray}
where $P_{ij}(\tau)$ is the probability that an electron  
in dot $j$ is found in dot $i$ after time $\tau$. In Eq.\
(\ref{eq:cij0semi}) we canceled a factor $2 \pi \hbar \nu_j$ in the
denominator against the phase space volume of the $j$th quantum dot.
For a quantum dot network, $P_{ij}(\tau)$ can be expressed in terms of 
a rate matrix $\tilde \gamma$,
\begin{equation}
  P_{ij}(\tau) = (e^{-\tilde \gamma \tau})_{ij},\ \
  \tilde \gamma = \tilde g/(2 \pi \hbar \tilde \nu).
  \label{eq:Ptau}
\end{equation}
Integrating over time, we then find
\begin{equation}
  \tilde c_{ij} =
  \tilde g^{-1}_{ij}.
\end{equation}

The interference between a path $\alpha$ and its time-reversed is 
suppressed if time-reversal symmetry 
is broken by a magnetic field, because a magnetic field changes the
phases of $A_{\alpha}$ and $A_{\bar \alpha}$ in opposite
ways.
Interference 
is also suppressed because of electron-electron interactions at a
finite temperature. Interactions cause the electrons to experience a
time-dependent potential $\phi(\vec{r},t)$, which modifies the phase 
of $A_{\alpha}$ and $A_{\bar \alpha}$ in different ways if the
trajectories $\alpha$ and $\bar \alpha$ are in different dots at the
same time $t$.\cite{altshuler1982b}
For a network
of quantum dots, the fluctuating potential $\phi$ is uniform inside
each dot, so that we can write $\phi(j,t)$, where $j=1,\ldots,\ND$ is 
the index
labeling the quantum dots in the network. For each amplitude $A_{\alpha}$
one then has\cite{altshuler1982b}
\begin{equation}
  A_{\alpha}[\phi] \to A_{\alpha}[0] e^{i \int_0^{t_{\alpha}}
  \phi(j_{\alpha}(t),t)/\hbar},
  \label{eq:Aphi}
\end{equation}
where $t_{\alpha}$ is the duration of the path $\alpha$,
$j_{\alpha}(t)$ the index of the quantum dot corresponding to the
position of path $\alpha$ at time $t$, and $A_{\alpha}[0]$ the return
amplitude in the absence of the potential $\phi$. 

For a quantum dot network, one may consider $\phi$ as a classical
fluctuating potential. (This will be verified in the exact quantum
mechanical calculation of Sec.\ \ref{sec:rmt} below.)
Its fluctuations are given by the fluctuation-dissipation
relation,\cite{landau1958}
\begin{eqnarray}
  \label{eq:phifluct1}
  \langle \phi (i,t) \phi(j,t') \rangle  
  & =  &  \int \frac{d \w }{ 2 \pi} e^{ - i \w ( t - t')/\hbar} 
  \frac{ 2 T}{ \w} 
  \Ima \left[ \LR_{ij} ( \w) \right],
  \nonumber \\
\end{eqnarray}
where the response function $\LR_{ij}(\w)$ describes the (linear)
change $ \delta \phi_i / e $ of the electric potential in the $i$th quantum 
dot to a change $\delta q_j = e \delta n_j$ of the
charge in the $j$th quantum dot,
\begin{eqnarray}
  \delta \phi_i(\omega) = 
  - L_{ij}^{\rm R}(\omega) \delta n_{j}(\omega).\ \
\end{eqnarray}
For the quantum dot network, one has
\begin{equation}
  L_{ij}^{\rm R}(\omega) = 
  - \left[ \tilde C/e^2 +
  d_{\rm s}  (\tilde \nu^{-1} - 2 \pi i \w \tilde g^{-1})^{-1} \right]^{-1}_{ij},
  \label{eq:LRdef}
\end{equation}
where the matrices $\tilde C$, $\tilde \nu$, and $\tilde g$ were defined in
Sec.\ \ref{sec:network} above. Typically, $\tilde C_{ii}/e^2 \ll
\nu_i$, $\tilde g_{ii}/|\omega|$, and we can replace Eq.\
(\ref{eq:LRdef}) by
\begin{equation}
  L_{ij}^{\rm R}(\omega) =
  \frac{1}{ d_{\rm s}} (2 \pi i \omega \tilde g^{-1} - \tilde \nu^{-1})_{ij}. \ \
  \label{eq:LRapprox}
\end{equation}
Using this expression for $\LR_{ij}(\omega)$, we find that Eq.\
  (\ref{eq:phifluct1}) simplifies to
\begin{eqnarray}
  \langle \phi (i,t) \phi(j,t') \rangle  
  & = & 
  \frac{ 4 \pi \hbar T}{ d_{\rm s}} \,  \tilde{g}^{-1}_{ij} \, \delta ( t - t'). 
  \label{eq:phifluct}
\end{eqnarray}

\begin{figure}
  \includegraphics{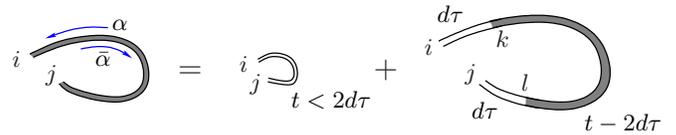}
  \caption{\label{fg:dyson}
  Calculation of the Cooperon propagator for a network of quantum
  dots. A trajectory $\alpha$ originating in dot $j$ and ending in dot
  $i$ and duration $t$
  is separated into two segments of duration $d\tau$ and a remaining
  segment of duration $t - 2 d\tau$ if $2 d \tau < t$. A self-consistent
  equation for $\tilde c_{ij}$ is obtained by considering the combined effect
  of escape, the magnetic field, and the fluctuating potential to first
  order in $d\tau$.}
\end{figure}

In order to find the effect of the fluctuating potential on the
Cooperon propagator $\tilde c_{ij}$, we separate the contributions from
trajectories $\alpha$ of duration $t_{\alpha}$ smaller and larger than
$2 d\tau$, where $d\tau$ is a time interval sufficiently short that
the net phase shift from the fluctuating potential in the exponent 
in Eq.\ (\ref{eq:Aphi}) is small, see Fig.\ \ref{fg:dyson}. 
We also take $d\tau$ much shorter 
than the dwell time in a single quantum dot, so that $P_{ij}(d\tau) =
\delta_{ij} - \tilde \gamma_{ij} d\tau$, see Eq.\ (\ref{eq:Ptau})
above. For trajectories of duration $t_{\alpha} > 2 d\tau$ we 
consider the initial and final segment of duration $d\tau$ 
separately. Recognizing that the contribution from the intermediate
segments of duration $t_{\alpha} - 2 d\tau$ can again be expressed in
terms of $\tilde c$, and using Eq.\ (\ref{eq:phifluct}) to average
over the fluctuating potentials, we then find 
\begin{eqnarray}
  \tilde c_{ij} &=&
  \frac{2 d\tau}{2 \pi \hbar \nu_i}
  \delta_{ij}
  \nonumber \\ && \mbox{}
  +
  \sum_{k,l=1}^{\ND}
  (\delta_{ik} - \tilde \gamma_{ki} d\tau)
  (\delta_{jl} - \tilde \gamma_{lj} d\tau)
  \tilde c_{kl}
    \nonumber \\ && \mbox{}
  -   \sum_{k,l=1}^{\ND}
  (\tilde \gamma_{{\rm H},ik} +
  \tilde \gamma_{{\rm H},jl} +
  \tilde \gamma_{\phi,ij}) \delta_{ik} \delta_{jl}
  \tilde c_{kl}d\tau,
  \nonumber \\ &=&
  \tilde c_{ij}
  + \frac{d\tau}{\pi \hbar \nu_i} \delta_{ij}
  - ( \Gamma + \Gamma_{\rm H} + \Gamma_{\phi} )_{ik,jl} \tilde c_{kl} d\tau,
\end{eqnarray}
up to corrections of order $d\tau^2$. Here
\begin{eqnarray}
  \tilde \gamma_{{\rm H},ij} &=& \frac{g_{{\rm H},i}}{2 \pi \hbar
  \nu_i} \delta_{ij}, \nonumber \\
  \tilde \gamma_{\phi,ij} &=& \frac{4 \pi T}{d_{\rm s} \hbar}
  (\tilde g^{-1}_{ii} + \tilde g^{-1}_{jj} - 2 \tilde g^{-1}_{ij}), 
\end{eqnarray}
and  
$ \Gamma_{ik,jl} = \tilde \gamma_{ki} \delta_{jl} 
+ \delta_{ik} \tilde \gamma_{lj} $, $(\Gamma_{\rm H})_{ik,jl} = 
\tilde \gamma_{{\rm H},ik} \delta_{jl} + \delta_{ik} \tilde \gamma_{{\rm H},jl} $,
 $ (\Gamma_{\phi} )_{ik,jl} = \tilde \gamma_{\phi,ij} \delta_{ik} \delta_{jl}$, 
cf. Eq.\ (\ref{eq:Gamma}) above.
Solving this equation for $\tilde c$, we arrive at Eq.\
(\ref{eq:cijdef}) of the previous section.

It is worth while to point out that the temperature dependence of 
weak localization is caused by processes that involve the
exchange of energy quanta small in comparison to the temperature. Such
processes are commonly referred to as ``dephasing'', in contrast to
more general inelastic processes which lead to a broadening of the
electronic distribution function.\cite{altshuler1985a,imry2002} 
In this sense, interaction effects in
the quantum dot network differ from those in a single quantum dot,
where weak localization is suppressed by inelastic processes that
involve a large energy transfer.\cite{sivan1994,blanter1996}
Indeed, the characteristic energy
exchanged in the electron-electron interactions scales with the
inverse of the dwell time $\hbar/\tau_{\rm D}$ in each quantum dot 
--- an observation that is closely related to the uniformity of the
interaction potential inside a quantum dot. The number of quanta 
exchanged along a typical trajectory is too small to lead to a 
significant broadening of the distribution function --- in that sense
transport in a quantum dot network is always quasi-elastic ---, 
although the exchange of a single quantum is sufficient to suppress 
the interference from time-reversed trajectories.

The semiclassical arguments of this section relied on the 
treatment of $\phi(\vec{r},t)$ as
a classical fluctuating potential. In this respect, we follow earlier
works on quantum dots by Seelig and B\"uttiker\cite{seelig2001} and on
granular metals by Blanter {\em et al.}\cite{blanter2006} This
approach was taken originally by Altshuler {\em et al.}\ for
dephasing in quasi one-dimensional and two-dimensional disordered
metals.\cite{altshuler1982b} In the next section, we confirm the
validity of this approach in the present context
by performing a fully quantum mechanical
calculation of the weak localization correction to first order in the
interaction propagator $L$. The calculation of Sec.\ \ref{sec:model}
shows that the potential fluctuations are essentially classical if $T
\gtrsim \hbar/\tau_{\rm D}$, where $\tau_{\rm D}$ is the (typical)
dwell time in a quantum dot in the network. Since $\hbar/\tau_{\rm D}$
is much smaller than the relevant temperature scale $T_{\phi}$ for the 
suppression of the weak localization correction by electron-electron 
interactions, cf.\ Eq.\ (\ref{eq:Tphi}) of Sec.\ \ref{sec:network}, 
this proves the
validity of our approach for all temperatures of interest.

\section{Quantum mechanical calculation}
\label{sec:model}

\subsection{Random matrix formulation}

We consider a network of $\ND $ chaotic quantum dots coupled to electron
reservoirs. The Hamiltonian of the entire system is written as
\begin{equation}
  \hat H = \hat H_0 + \hat H_{\rm int},
\end{equation}
where $\hat H_0$ describes the electrons inside the quantum dots or
inside leads without taking into account their interactions, and $\hat
H_{\rm int}$ describes the electron-electron interactions. We write the
non-interacting Hamiltonian $\hat H_0$ as a sum of three terms,
\begin{equation}
  \hat H_0 = \hat H_{\rm D} + \hat H_{\rm DL} + \hat
  H_{\rm L},
\end{equation}
where $\hat H_{\rm D}$ and $\hat H_{\rm L}$ describe the electrons
inside the quantum dot network and inside the leads, respectively, whereas
$\hat H_{\rm DL}$ describes the coupling between the quantum dots and
the leads. We now describe each of the three terms contributing to
$\hat H$ separately.

Linearizing the electronic spectrum around the Fermi energy inside the leads, we have
\begin{eqnarray}
\hat{H}_{\rm L} &=& 
  \sum_{a=1,2}
  \sum_{j=1}^{N_{a}}\int \frac{dk}{2\pi}\, 
  \vF_{a,j} k \, \psd_{a,j}(k)\ps_{a,j}(k),
  \label{eq:Hl}
\end{eqnarray}
where the index $a=1,2$ labels leads connecting to the
left and right electron reservoirs. The operators $\psd_{a,j}(k)$
and $\ps_{a,j}(k)$ are for electrons in scattering states at
wavenumber $k$ (measured with respect to the Fermi wavenumber) and
transverse mode $j$. The total number of propagating modes in the
leads connecting to reservoir $a$ is $N_{a}$, $a=1,2$. [If a reservoir
 is coupled to more than one lead, the summation 
over the index $j$ represents a sum over the transverse 
modes in all leads connected to the given reservoir.] 
Finally, $v_{a,j}$ is the Fermi velocity 
of electrons in mode $j$. The current operator $\hat
I_{a}$ reads
\begin{equation}
  \hat I_{a} =
  e \sum_{j=1}^{N_a}
  v_{a,j} 
  \left(
  \psd_{a,j+}
  \ps_{a,j+} 
  - \psd_{a,j-}
  \ps_{a,j-}
  \right),\ \ {a = 1,2},
  \label{eq:cur}
\end{equation}
where 
\begin{equation}
  \ps_{a,j\pm} =
  \int \frac{dk}{2 \pi}
  e^{\pm i k \delta} \ps_{a,j}(k),\ \ {a = 1,2} ,
\end{equation}
and $\delta  > 0 $ is a positive infinitesimal.

We use random matrix theory to describe the quantum dots. Following
standard procedures, the electron operators in each quantum dot are 
represented by an $M_j$-component vector $\ps_j$, where the
index $j=1,\ldots,\ND $ labels the quantum dots in the
network and $M_j$ is the dimension of the subspace corresponding to
the dot with index $j$. 
The Hamiltonian $\hat H_{\rm D}$ then reads
\begin{eqnarray}
\hat{H}_{\rm D} &=& \sum_{i=1}^{\ND}
  \sum_{\alpha,\beta=1}^{M_i}
  \psd_{i,\alpha}
  H_{i,\alpha \beta}^{\phantom{\dagger}} \ps_{i,\beta}
  \nonumber \\ && \mbox{} +
  \sum_{i < j} \sum_{\alpha, \beta} 
  \left(  \, \psd_{i,\al}  V_{ij,\ab}^{\phantom{\dagger}}
  \ps_{j,\be} + \mbox{h.c.} \right).
  \label{eq:Hdd} \label{eq:Hd}
\end{eqnarray}
Here the elements $H_{i,\alpha\beta}$ of the $M_i$-dimensional
matrices $H_i$ are random numbers taken from from a Gaussian
distribution with zero mean and with variance 
\begin{eqnarray}
  \langle H_{i,\alpha \beta} H_{i,\gamma \delta} \rangle
  &=&
\langle H_{i,\alpha \beta} H_{i,\delta \gamma}^* \rangle 
  \nonumber \\ & = & \frac{\lambda_{i}}{M_i} 
  \delta_{\alpha \delta} \delta_{\beta \gamma} + 
  \frac{\lambda_i'}{M_i}  \delta_{\alpha \gamma} \delta_{\beta
  \delta}.
  \label{eq:Havg}
\end{eqnarray}
The parameters $\lambda_i$ and $\lambda_i'$ are related to the density 
of states $\nu_i$ and magnetic flux $\Phi_i$ in each quantum
dot,\cite{aleiner2002} $i=1,\ldots,\ND$,
\begin{equation}
  \lambda_i = \frac{M_i^2}{\pi^2 \nu_i^2},\ \
  \lambda_i' = \frac{M_i^2}{\pi^2 \nu_i^2}
  \left(1 - \frac{E_{{\rm Th},i} \nu_i \Phi_i^2}{4 M_i \Phi_0^2} \right),
\end{equation}
where $\Phi_0$ the flux quantum and $E_{{\rm Th},i}$ is the Thouless
energy of the $i$th quantum dot. 
Further, in Eq.\ (\ref{eq:Hd}), 
the $M_i \times M_j$ matrices $V_{ij}$ are related to the
transmission matrices $t_{ij}$ of the contact between dots $i$ 
and $j$,
\begin{equation}
  t_{ij} = 2 \pi V_{ij} (\nu_i \nu_j M_i M_j )^{1/2}
  (M_i M_j + \pi^2 \nu_i \nu_j V_{ij}^{\dagger} V_{ij})^{-1}.
  \label{eq:tij}
\end{equation}
The Hamiltonian $H_{\rm DL}$ describing the coupling between the dots
and the leads reads
\begin{eqnarray}
\hat{H}_{\rm DL} & = & 
  \sum_{a=1}^{2}
  \sum_{j=1}^{N_{a}}
  \sum_{i=1}^{\ND} \sum_{\alpha=1}^{M_i} 
  \int \frac{dk}{2\pi} 
  \nonumber \\ && \mbox{} \times
  \left( \psd_{i,\alpha} W_{ia,\al j }^{\phantom{\dagger}} 
  \ps_{a,j}(k) + 
  {\rm h.c.}
  \right), \nonumber \\
 \label{eq:Hdl}
\end{eqnarray}
where the $N_i \times N_a$ matrices $W_{ia}=W_{ai}^{\dagger}$ 
are related to the transmission matrices $t_{ia}$ of the
contact between the $i$th quantum dot and reservoir $a$,
\begin{equation}
  t_{i a} = 2 \pi W_{ia} (\nu_{a} \nu_i M_i)^{1/2}
  (M_i + \pi^2 \nu_i \nu_{ a}^{1/2} W_{a i} W_{i{
  a}} \nu_{ a}^{1/2})^{-1},
\end{equation}
with $a=1,2$ and $\nu_{a}$ is an $N_{
  a}$-dimensional matrix with elements $(\nu_{ a})_{ij}
=\delta_{ij} (2 \pi \hbar \vF_{a,j})^{-1}$. 
The dimensionless conductance $g_{ij}$ and and form factor $f_{ij}$
of the contact between dots $i$ and $j$ are defined in terms of the
transmission matrix $t_{ij}$ as in Eq.\ (\ref{eq:fgdef}). Similarly,
the dimensionless conductance $g_{ia}' = g_{ai}'$ and form factor
$f_{ia}' = f_{ai}'$ between the dots and the two electron reservoirs
are defined in terms of $t_{ia}'$ as in Eq.\ (\ref{eq:fgpdef}).

For the electron-electron interaction we take density fluctuations
inside each dot to be well screened, so that the interaction couples
to the total charges of the dots only,
\begin{equation}
\hat{H}_{\rm int} \, = \, \sum_{i,j} \frac{e^2}{2} \hat{n}_{\rm i} \left[
  \tilde{C}^{-1} \right]_{ij} \hat{n}_{\rm j},\ \
  \hat n_i = \sum_{\alpha=1}^{M_i} \hat \psi^{\dagger}_{i,\alpha}
  \hat \psi^{\vphantom{M}}_{i,\alpha},
\end{equation}
where the capacitance matrix $\tilde C$ was defined in Eq.\
(\ref{eq:Cdef}) above.
The corresponding interaction Hamiltonian for a single quantum
 dot is known as `universal interaction Hamiltonian'.\cite{aleiner2002}

Evaluating the conductance $g$ of the quantum dot network and its 
leading interaction corrections using the Kubo formula one finds
\begin{equation}
  G = \frac{ d_{\rm s} e^2}{h} g,\ \
  g = g_0 + \delta g^{\rm deph} + \delta g^{\rm  int},
  \label{eq:gsep}
\end{equation}
where $g_0$ is the conductance in the absence of interactions ({\em  i.e.,} 
for Hamiltonian $\hat H_0$), and $\delta g^{\rm deph}$ and $\delta
g^{\rm  int}$ are interaction corrections. 
(The reason for the separation between $\delta g^{\rm deph}$ and
$\delta g^{\rm int}$ is that these two corrections have different
temperature dependences, as will become apparent later.)
Denoting with ``$\cdot$'' adjacent indices to be summed over [as
in Eq.\ (\ref{eq:gcl})], the three terms in Eq.\ (\ref{eq:gsep}) read
\begin{widetext}
\begin{equation}
  g_{0} = 4 \pi^2 \int d \e  
  \left[-   \partial_{\e} f(\varepsilon) \right]
  \mbox{tr}\, \nu_1 W_{1\cdot}
  \GR_{\cdot\cdot}(\e) W_{\cdot 2} \nu_2 W_{2\cdot}
  \GA_{\cdot\cdot}(\e) W_{\cdot 1}, \label{eq:Gni}
\end{equation}
and the interaction corrections $\delta g^{\rm deph}$ and $\delta g^{\rm  int}$ are
\begin{eqnarray}
\delta g^{\rm deph} & = 
&  4 \pi^2  \int d\e \int
  \frac{d \w}{ 2\pi } 
  \left[ - \partial_{\e} f(\varepsilon) \right]
  \left[ \coth(\w / 2T) + \tanh(( \e - \w)/2T) \right] \,  
  \sum_{i,j=1}^{\ND}
   \Ima \left[ \LR_{ij}(\w ) \right]  \nonumber \\
  && 
  \mbox{} \times 
  \mbox{tr}\,
  \Big[ \nu_1
  W_{1\cdot} \GR_{\cdot i}(\e) \GR_{ij}(\e - \w) \GR_{j\cdot}(\e) 
  W_{\cdot 2} \nu_2
  W_{2 \cdot} \GA_{\cdot\cdot}(\e) W_{\cdot 1}  
  \nonumber \\ && \ \ \ \mbox{} + 
  \nu_1 W_{1\cdot} \GR_{\cdot\cdot}(\e) W_{\cdot 2} \nu_2
  W_{2 \cdot} \GA_{\cdot i}(\e) \GA_{ij}(\e - \w) \GR_{j \cdot}(\e)
  W_{\cdot 1}  
  \nonumber \\ && \ \ \ \mbox{} + 
  \frac{1}{2} \nu_1 W_{1\cdot} \GR_{\cdot i}(\e - \w) \GR_{i \cdot}(\e) W_{\cdot
    2} \nu_2
  W_{2 \cdot} \GA_{\cdot j}(\e)  \GA_{j \cdot}(\e - \w) W_{\cdot 1} 
  \nonumber \\ && \ \ \ \mbox{} + 
  \frac{1}{2} \nu_1 W_{1\cdot} \GR_{\cdot i}(\e ) \GR_{i \cdot}(\e - \w) W_{\cdot
    2} \nu_2
  W_{2 \cdot} \GA_{\cdot j}(\e - \w )  \GA_{j \cdot}(\e) W_{\cdot 1} 
  \Big]  \label{eq:Gdeph}
\end{eqnarray}
\begin{eqnarray}
  \delta g^{\rm int} & = 
  & 4 \pi^2 \int d \e \int
  \frac{d \w}{2 \pi } 
  \left[ - \partial_{\e} f(\varepsilon) \right]
  \tanh((\e - \w)/2T)\,  
  \sum_{i,j=1}^{\ND}
  \Ima  \Big[ \LA_{ij}(\w)
  \nonumber \\ && \mbox{} \times
  \mbox{tr}\,  \left[ 
  \nu_1 W_{1\cdot} \GR_{\cdot i}(\e) 
  \GR_{ij}(\e - \w) \GR_{j \cdot}(\e) W_{\cdot 2} \nu_2
  W_{2 \cdot} \GA_{\cdot\cdot}(\e) W_{\cdot 1}
  \right. \nonumber \\ && \left. \ \ \
  \mbox{} + \nu_1
  W_{1\cdot} \GR_{\cdot\cdot}(\e) W_{\cdot 2} \nu_2
  W_{2 \cdot} \GA_{\cdot i}(\e) \GR_{ij}(\e - \w) \GA_{j\cdot}(\e)
  W_{\cdot 1} \right] \Big]. \label{eq:Gint}
\end{eqnarray}
\end{widetext}
In these equations $\GR_{ij}$ and $\GA_{ij}$ denote the retarded and
advanced Green functions of the network of quantum dots without the
electron-electron interaction Hamiltonian $\hat{H}_{\rm int}$. These are matrices of
dimension $M_i \times M_j$, which are the solution of
\begin{eqnarray}
  \left[\e - H_{i} + i \pi \sum_{a=1}^{2} W_{ia} \nu_a W_{ai} \right]
  \GR_{ii}(\e)
  + V_{i\cdot} \GR_{\cdot i}(\varepsilon) &=& \openone_i, 
  \nonumber \\
  \left[\e - H_{i} - i \pi \sum_{a=1}^{2} W_{ia} \nu_a W_{ai} \right]
  \GA_{ii}(\e)
  + V_{i\cdot} \GA_{\cdot i}(\varepsilon) &=& \openone_i, 
  \nonumber \\
  \label{eq:GReq}
\end{eqnarray}
with $\openone_i$ the $M_i \times M_i$ unit matrix. Finally,
$\LR_{ij}(\w)$ and $\LA_{ij}(\w) = \LR_{ij}(\w)^*$ represent the 
(RPA) screened interaction propagator, see Eq.\ (\ref{eq:LRdef})
above.

It remains to calculate the ensemble average of the conductance $G$
for the ensemble of Hamiltonians described by Eq.\ (\ref{eq:Havg})
above. This is the subject of the next subsection.

\subsection{Average over random Hamiltonian}
\label{sec:rmt}

The average over the random matrices $H_i$ is performed using a
variation of the impurity diagrammatic technique.\cite{zee2003} 
This technique has been applied for various transport and thermodynamic 
properties of chaotic quantum dots without electron-electron 
interactions.\cite{melsen1996,vavilov1999,clerk2000,vavilov2001a}
Below we present its generalization to arbitrary networks.

\begin{figure}
  \includegraphics{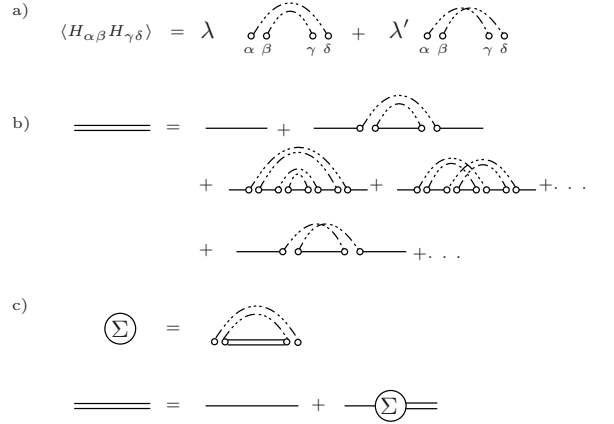}
  \caption{(a) Diagrammatic rules for the ensemble average using Random
  Matrix Theory. The weight factors depend on the symmetry present: 
  $\lambda' = \lambda $ in the presence of time reversal symmetry, while 
  $\lambda'$ is reduced in the presence of a weak magnetic field and 
  $\lambda' = 0 $ where 
  time reversal symmetry is fully broken. 
(b) Expansion of the full matrix propagator in terms of single
  propagators $1/( \e + i \pi \nu WW^{\dagger} )$, depicted by single
  lines, and the matrix elements $H_{\alpha \beta}$, depicted by two
  open circles. (c) Dyson equation for the self energy $\Sigma$. \label{fg:GetS}}
\end{figure}

\subsubsection{Average Green function}

We first discuss the calculation of the ensemble average of the Green
function, $\langle \GR_{ij}(\e) \rangle$ and $\langle \GA_{ij}(\e)
\rangle$. Following the diagrammatic rules laid out in Fig.\
\ref{fg:GetS} and keeping diagrams in the 
non-crossing approximation only,\cite{Abrikosov1963}
{\em i.e.} diagrams without crossing double lines, one
finds that the ensemble averaged Green function $\langle
\GR_{ij}(\e) \rangle$ satisfies the Dyson equation
\begin{equation}
  \langle \GR_{ij}(\e) \rangle =
  \GR_0(\e)_{ij} + \sum_{k} \GR_0(\e)_{ik} \Sigma_k
  \langle \GR_{kj}(\e) \rangle , \label{eq:GR1}
\end{equation}
where the self energy $\Sigma_k$ is 
\begin{equation}
  \Sigma^{\rm R}_k(\e) = 
  \frac{\lambda_k}{M_k} \mbox{tr}\, \langle  \GR_{kk}(\e) \rangle,
  \label{eq:GR2}
\end{equation}
and $\GR_0(\e)$ is the solution of Eq.\ (\ref{eq:GReq}) with $H_{i} 
= 0$. Combining Eqs.\ (\ref{eq:GR1}) and (\ref{eq:GR2}) gives a 
self-consistent equation for $\Sigma^{\rm R}$. In the limit $M_i \gg
  g_{i1}' + g_{i2}' + \sum_{j \ne i} g_{ij}$, one finds
\begin{eqnarray}
  \langle \GR_{ij}(\e) \rangle &=& 
  \langle \GA_{ji}(\e) \rangle^{\dagger}
  \nonumber \\ 
& = &  - \frac{i \pi }{ M_i + \Delta_i} \tilde{\nu}_{ij}
  - \sqrt{ \frac{\pi^2 \nu_i \nu_j}{4 M_i M_j}} t_{ij}
  \nonumber \\ && \mbox{}
  + \frac{\pi  }{2 M_i^2}
  \left( \pi \nu_i \e - i
  \mbox{tr}\, \frac{\Delta_i}{M_i + \Delta_i}
  \right) \tilde{\nu}_{ij},
\end{eqnarray}
where $\tilde{\nu}_{ij} $ and  $t_{ij}$ are  given in Eq.\ (\ref{eq:tildenu}) and (\ref{eq:tij}) above and
$\Delta_i$ is an hermitian $M_i \times M_i$ matrix,
\begin{equation}
  \Delta_i \, = \,  
  \pi^2 \nu_i \sum_{k \neq i}^{N_{\rm D}}
  \frac{1}{M_k} V_{ik} \nu_k V_{ki} 
  + \pi^2 \nu_i \sum_{a=1}^{2} W_{ia} \nu_a W_{ai}.
  \label{eq:Deltaidef}
\end{equation}
 
\subsubsection{Classical conductance}

\begin{figure}
  \includegraphics{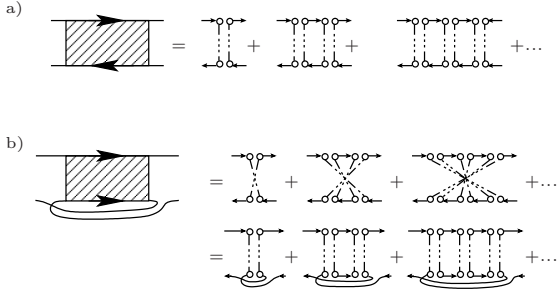}
  \caption{Diffuson ladder (a) and Cooperon ladder (b). 
  \label{fg:ladders}}
\end{figure}

To leading order in the average number $N$ of transmitting channels per dot,
 the calculation of the average conductance
involves the calculation of geometric series involving the ensemble
averaged Green functions. Diagrammatically, these geometric
series correspond to ``ladder diagrams'', as shown in Fig.\
\ref{fg:ladders}. Such ladders are the equivalent of the ``diffuson'' 
propagator in diagrammatic perturbation theory. The building block of
the geometric series is
\begin{eqnarray}
\tr \langle \GR_{ij}(\e) \rangle \langle \GA_{ji}(\e') \rangle & = 
&  
  \frac{\pi^2 \nu_i^2}{\Mi} \delta_{ij}
  \\ && \mbox{}
  - \frac{\pi^2 \nu_i \nu_j}{4 M_i M_j}
  \left[\tilde g -  i 2\pi(\e-\e') \tilde{\nu} \right]_{ij},
  \nonumber
\end{eqnarray}
where $\tilde g_{ij}$ was defined in Eq.\ (\ref{eq:tildeg})
above. Summing the geometric series in Fig.\ \ref{fg:ladders}a then gives the Diffuson matrix
\begin{eqnarray}
  D_{ij}(\e,\e') &=&
  \frac{2 M_i}{\pi \nu_i}
  \left[ \tilde g - i 2 \pi  (\e - \e') \tilde{\nu} \right]^{-1}_{ij}
  \frac{2 M_j}{\pi \nu_j}.
  \label{eq:Diff}
\end{eqnarray}

For the calculation of the mean conductance one also needs a trace
that involves the lead indices,
\begin{eqnarray}
  D_{ia}' &=&
  \pi \nu_a
  \mbox{tr}\,  \left[W_{ai} \langle \GR_{ii} \rangle
  \langle  \GA_{ii}  \rangle W_{ia} \right]
  \nonumber \\
  & = & \pi \nu_i \frac{ g_{ai}'}{4 M_i},\ \
  a = 1,2.
  \label{eq:Lia}
\end{eqnarray}
Combining everything, we then find the leading conductance of the system  
$$
  \langle g \rangle = 
   g_{1 \cdot}' (\tilde g^{-1})_{\cdot \cdot} g_{\cdot 2}',
$$
which is equation (\ref{eq:gcl}) of section \ref{sec:network}.

\begin{figure}
  \includegraphics{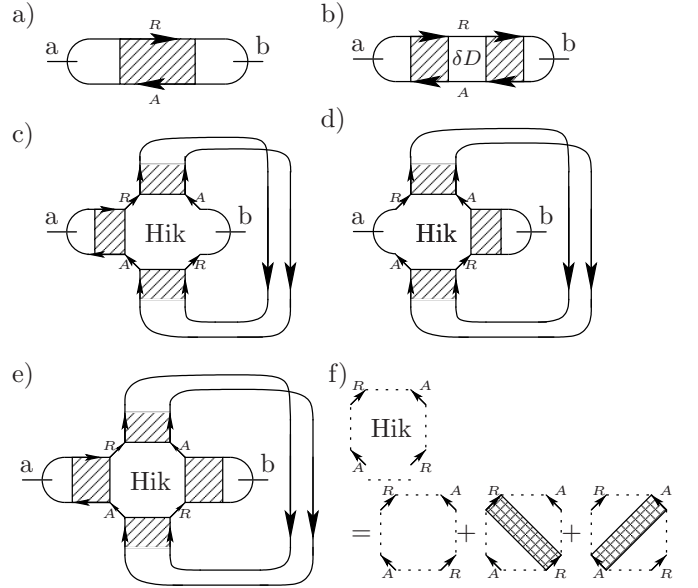}
  \caption{(a) Diagrammatic representation of the leading contribution
  $g^{\rm cl}$ to the ensemble-averaged conductance $\langle g
  \rangle$. (b)--(e) Diagrams contributing to the weak localization
  correction $\delta g^{\rm WL}$. (f) Definition of the Hikami-box used in (c)--(e).\label{fg:Gni_ext}}
\end{figure}

\subsubsection{Weak localization correction}

The above calculation gives the conductance to leading order in
$g$. A correction to sub-leading order in $g$ is given by a class of
diagrams that contains a maximally crossed ladder, as shown in Fig.\
\ref{fg:ladders}b. These contributions are analogous to the ``Cooperon''
contributions in diagrammatic perturbation theory.\cite{altshuler1985a}
 The summation of the geometric series promotes
the contribution to be of order $1/N$ instead of $1/M$, as is the
naive expectations for diagrams that contain one crossed line.

\begin{figure}
  \includegraphics{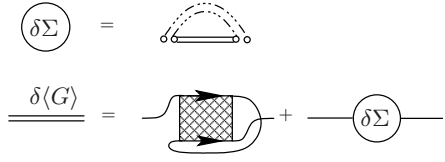}
  \caption{ Dyson equation for corrections to $\langle G_{ii}\rangle $ 
  due to the possibility of Cooperon like ladders in the time 
  reversal symmetric case. Double-hatching indicates a retarded-retarded
   or advanced-advanced pairing. These ladders are parametrically small,
   and for that reason can also not extend across multiple dots.
   \label{fg:dG_C}}
\end{figure}

In contrast to the diffuson propagator discussed above, the cooperon
propagator is sensitive to magnetic flux. Proceeding as before, we
find
\begin{equation}
  C_{ij}(\e , \e') \, = \,   
  \frac{2 M_i}{\pi \nu_i}
  \left[ \tilde g_{\rm H} +
  \tilde g - 2 \pi i \tilde \nu (\e - \e') \right]^{-1}_{ij}
  \frac{2 M_j}{\pi \nu_j},
\end{equation}
with $g_{\rm H}$ defined in Eq.\ (\ref{eq:gH}). For the calculations below, we also need geometric series of 
Green functions of the same type. These read 
\begin{eqnarray}
  C_{ij}^{\rm RR} (\e,\e') 
  & = & C_{ij}^{\rm AA}(\e, \e')^* 
  \nonumber \\ 
  & = & \frac{1}{16\pi^2 \nu_i \nu_j} \Big( 
  \big( 8 M_i + \tilde g_{{\rm H}, ii}+ \tilde{g}_{ii} 
  \nonumber \\ && \mbox{} 
  - i 2 \pi (\e + \e') \nu_i \big)\delta_{ij} 
  -  \tilde{g}_{ij} \left( 1  - \delta_{ij}\right ) \Big).
 ~~~~ \label{eq:Coop}
\end{eqnarray}
 
Cooperon ladders give a correction to the self-energy appearing in the
calculation of the average Green function, as depicted in Fig.\
\ref{fg:dG_C}. Calculation of the self-energy correction $\delta
\Sigma_i$ to leading order in $g/M$ then gives
\begin{eqnarray}
  \delta \Sigma_i & = &
  \frac{\lambda_i}{M_i} \tr \left[ 
  \langle \GR_{ii} \rangle ( C^{\rm RR}_{ii} \langle \GR_{ii} \rangle + 
  \delta \Sigma_i ) \langle \GR_{ii} \rangle \right] \nonumber \\
& = &  \frac{i}{4 \pi \nu_i} \mbox{.}
\end{eqnarray}
As this contribution is already small as $1/M$, one may neglect the
effect of a weak magnetic field on this term. The self energy
correction $\delta \Sigma$ affects the diffusion ladders as $D \to D +
\delta D$, with 
\begin{equation}
  \delta D_{ij} = - \frac{\pi^2 \nu_i^2}{2 M_i^2} \delta_{ij} \mbox{.}
\end{equation}
This contribution is depicted in figure \ref{fg:Gni_ext}b.

In the diagrams for the weak localization correction to the
conductance, the Cooperon and diffuson propagators are connected in a
so-called ``Hikami box''.\cite{Hikami1981} In our diagrammatic analysis the analogue of
a Hikami box is depicted in figure \ref{fg:Gni_ext}f. 
We consider the general case of a Hikami box with four energy
arguments. We write $\e_1$ ($\e_1'$) for the energy argument of the 
retarded (advanced) 
matrix propagator on the left side, and $\e_2$ ($\e_2'$) for 
the energy argument 
of the retarded (advanced) propagator on the right.
 For the calculation of the weak localization 
correction one only needs the case of equal arguments, 
$\e_1 = \e_1' = \e_2 = \e_2'$. 
For dephasing and interaction corrections, some arguments
differ. Explicit
calculation shows that the Hikami box depends on the
combination
$\omega = \e_1'-\e_1+\e_2'-\e_2$ only. Hence we write
 $B_{ij,kl}(\omega)$, where the indices $i$
and $j$ refer to the left and right (Diffuson) ladders and the 
indices $k$ and $l$ refer to the bottom and top (Cooperon) ladders.

The calculation is essential but technical; we outline it in the appendix 
\ref{app:hik}. The Hikami box $B_{ij,kl}(\w)$ is zero except where at most two different indices appear, 
\begin{widetext}
\begin{eqnarray} 
  \label{eq:Hikami}
  B_{ij,kl}(\w) &=& \frac{\pi^4 \nu_i \nu_j \nu_k \nu_l}{16
  M_i M_j M_k M_l} 
  \left[ 2 \pi i \nu_i \w \delta_{ij} \delta_{jk} \delta_{kl} +
  (\delta_{il} \delta_{jk} + \delta_{ik} \delta_{jl})
  (\tilde g_{ij} + \tilde g_{{\rm H},ij} - \tilde f_{ij})
  \right. \nonumber \\ && \left. \mbox{}
  + (\delta_{ik} \delta_{il} + \delta_{jk} \delta_{jl})
  \tilde f_{ij}
  + (\delta_{ij} \delta_{ki} + \delta_{ij} \delta_{li}) \tilde f_{kl}
  - \delta_{ij} \delta_{kl} \tilde f_{ik}\right].
\end{eqnarray}

For the evaluation of the weak localization correction, one also needs
to consider Hikami boxes that are connected to the leads, not only to
Diffuson propagators inside the quantum dot network. The two 
contributions
of this type are depicted in figure \ref{fg:Gni_ext}c and d. 
They are
\begin{eqnarray}
  \label{eq:HikamiL}
  B'_{aj,jj} & = & B'_{ja,jj} \nonumber \\ &=& 
  -\frac{\pi^3 \nu_j^3}{ 16 M_j^3} f'_{aj}.
\end{eqnarray}

Combining everything, we have (see Fig.\ \ref{fg:Gni_ext})
\begin{eqnarray}
\label{eq:dGWLni}
\delta g^{\rm WL} &  = 
& 2 D_{1\cdot}' D_{\cdot\cdot} \delta
  D_{\cdot\cdot}
  D_{\cdot\cdot} D_{\cdot 2}' 
  +  2
  \sum_{i,j=1}^{\ND}
  C_{ij}
  \left( D_{1\cdot}' D_{\cdot\cdot}
  B'_{\cdot 2,ji}  
  + B'_{1\cdot,ji} D_{\cdot\cdot} D_{\cdot 2}'
  +   D_{1\cdot}' D_{\cdot\cdot}
  B_{\cdot\cdot,ji}(0) D_{\cdot\cdot} D_{\cdot 2}' \right),
\end{eqnarray}
where $D_{ia}' = D_{ai}'$ was defined in Eq.\ (\ref{eq:Lia}) above and we 
have suppressed superscripts as well as inconsequential energy arguments
  of $D^{\rm RA}(\e, \e) $, $C^{\rm RA}(\e,\e)$, cf.\ Eqs.\ (\ref{eq:Diff}), 
(\ref{eq:Coop}).
The four terms correspond to the four 
diagrams b - e of Fig.\ \ref{fg:Gni_ext}. 
Substituting our results for the Hikami box $B$, the Cooperon and
Diffuson propagators $C$ and $D$, and the interaction propagator
$L$, we arrive at Eq.\ (\ref{eq:dGWL}) of Sec.\ \ref{sec:network},
with the zero-temperature Cooperon
$\tilde c = (\tilde g + \tilde g_{\rm H})^{-1}$.

\begin{figure}
  \includegraphics{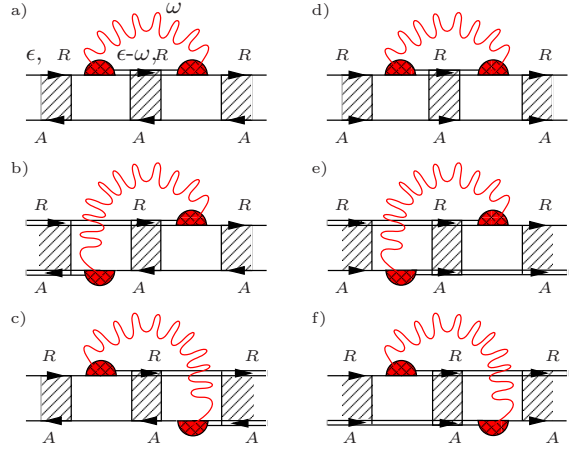}
\caption{Diagrams for the first-order dephasing correction. 
  Diagrams depicted in (b), (c) and (e), (f) are weighed 
  with a factor $1/2$, in line with Eq.\ (\ref{eq:Gdeph}).
  Together (a), (b) and (c) constitute the correction to 
  the Diffuson propagator, which cancels to leading order. 
  Hence the only relevant contributions are the corrections to the 
  Cooperon in (d), (e) and (f). In both cases, complex conjugate contributions 
  exist which are obtained by placing the vertices on the opposite
   matrix propagation lines. \label{fg:deph_gen}}
\end{figure}

So far we have not taken into account electron-electron
interactions. To lowest order in perturbation theory in the
interaction Hamiltonian $\hat{H}_{\rm int}$, 
the dominant interaction correction to weak localization
comes from $\delta g^{\rm deph}$ in
Eq.\ (\ref{eq:Gdeph}). The corresponding diagrams 
are depicted in Fig.\ \ref{fg:deph_gen}. We now calculate that correction.
This interaction correction is nonzero only if both interaction vertices
appear inside the Cooperon propagator. (This is why this interaction
correction does not affect the leading contribution $g_0$ to the
conductance.)

To calculate the interaction correction, one notices that the interaction
vertices are ``dressed'', as is shown in Fig.\ \ref{fg:vertren}. For
this case energy arguments may be neglected, as they lead to
corrections small in $g/M$. Labeling the dot in which the interaction
takes place by the index $\alpha$, the dressed interaction then reads
\begin{eqnarray} 
I^{\rm R}_{\al,ij} &=& (I^{\rm A}_{\al, ij})^{*} \nonumber \\ &=&
  \tr \left[ \langle \GR_{ii} \rangle \left( 1 + \tr \left[ \langle
  \GR_{ii} \rangle \langle \GR_{ii} \rangle \right]  D^{\rm RR}_{ii}
  \right)
  \langle \GR_{ii} \rangle \langle \GA_{ii} \rangle \right] 
  \delta_{\al i } \delta_{\al j}  \nonumber \\
  & = &  \frac{ \pi \nu_i}{2 M_i} 
  \left( - i 2 \pi \tilde{\nu}_{ij} \delta_{\alpha i} \right) 
  \frac{ \pi \nu_i}{2 M_i} 
\end{eqnarray}
where
\begin{eqnarray}
  D_{ij}^{\rm RR}(\e,\e') & = & D_{ij}^{\rm AA}(\e, \e')^*  
  \nonumber \\ 
& = & \frac{1}{16\pi^2 \nu_i \nu_j} \Big[ 
  \big( 8 M_i + \tilde{g}_{ii} - i 2 \pi (\e + \e') \nu_i \big)\delta_{ij}
  -  \tilde{g}_{ij} \left( 1  - \delta_{ij}\right ) \Big].
\end{eqnarray}
The interaction correction $\delta  C$ 
to the equal-energy Cooperon propagator
$\tC(\e,\e)$ then becomes
\begin{eqnarray}\label{eq:deph}
  \delta \tC_{ij}
  & = &
  \int d\e \int
  \frac{d \w}{ 2\pi } 
  \left[ - \partial_{\e} f(\varepsilon) \right]
  \left[ \coth(\w / 2T) + \tanh(( \e - \w)/2T) \right] \,  
  \sum_{\alpha,\beta=1}^{\ND}
  \mbox{Im}\, [\LR_{\alpha \beta}(\w)]
  \nonumber \\ && \mbox{} \times
  \left[
  \tC_{i\cdot}(\e,\e) I^{\rm R}_{\al,\cdot\cdot} 
  \tC_{\cdot\cdot}(\e - \w ,\e) I^{\rm R}_{\be,\cdot\cdot} \tC_{\cdot
    j}(\e,\e) 
  + \tC_{i\cdot}(\e - \w,\e) I^{\rm A}_{\al,\cdot\cdot} \tC_{\cdot\cdot}(\e - \w,\e- \w) I^{\rm R}_{\be,\cdot\cdot}
  \tC_{\cdot j}(\e , \e - \w) 
  + \mbox{c.c.}\right]. ~~~~
\end{eqnarray}
Performing the energy integration and passing to dimensionless propagators, we then find
\begin{eqnarray}
  \label{eq:deltacresult}
  \delta c_{ij}
  & = &    \int \frac{d\w}{2 \pi} 
  \frac{\w}{2 T \sinh^2 (\w/2 T)}
  \sum_{\alpha,\beta=1}^{\ND}
  \mbox{Im}\, [4 \pi^2 \nu_{\alpha} \nu_{\beta}
  \LR_{\alpha \beta}(\w)]
  \\ && \mbox{} \times
  \left\{\left(\tilde{g} + \tilde{g}_{\rm H}  + i 2 \pi \w \tilde{\nu} 
    \right)^{-1}_{i \al}
  \left(\tilde{g} + \tilde{g}_{\rm H} \right)^{-1}_{\al \be} 
  \left(\tilde{g} + \tilde{g}_{\rm H} -  i 2 \pi \w \tilde{\nu} 
    \right)^{-1}_{\be j} 
  - \left(\tilde{g} + \tilde{g}_{\rm H} \right)^{-1}_{i \al}
  \left(\tilde{g} + \tilde{g}_{\rm H} + i 2 \pi \w \tilde{\nu}  
    \right)^{-1}_{\al \be} 
  \left(\tilde{g} + \tilde{g}_{\rm H} \right)^{-1}_{\be j} 
  \right. \nonumber \\ && \left. \mbox{}
  + \left(\tilde{g} + \tilde{g}_{\rm H}  - i 2 \pi \w \tilde{\nu} 
    \right)^{-1}_{i \al}
  \left(\tilde{g} + \tilde{g}_{\rm H} \right)^{-1}_{\al \be} 
  \left(\tilde{g} + \tilde{g}_{\rm H} +  i 2 \pi \w \tilde{\nu} 
    \right)^{-1}_{\be j} 
  - \left(\tilde{g} + \tilde{g}_{\rm H} \right)^{-1}_{i \al}
  \left(\tilde{g} + \tilde{g}_{\rm H} - i 2 \pi \w \tilde{\nu}  
    \right)^{-1}_{\al \be} 
  \left(\tilde{g} + \tilde{g}_{\rm H} \right)^{-1}_{\be j} 
  \right\}. \nonumber
\end{eqnarray}

\begin{figure}
  \includegraphics{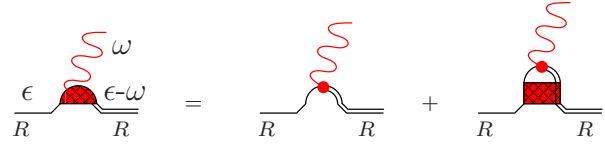}
  \caption{Renormalization of the interaction vertex by ladder diagrams
  involving Green's functions of the same type 
  (retarded-retarded or advanced-advanced).  \label{fg:vertren}}
\end{figure}

Let us now inspect the integral in Eq.\ (\ref{eq:deltacresult}). The
term between brackets $\{ \ldots \}$ is proportional to $\omega^{-2}$
if $\omega \gtrsim \hbar/\tau_{\rm D}$, where $\hbar/\tau_{\rm D} \sim
g/\nu$ is the inverse dwell time of a dot in the network. Since
$\mbox{Im}\, L^{\rm R}(\omega) \propto \omega$ for $\omega \sim
\hbar/\tau_{\rm D}$, one thus concludes that the integral in Eq.\
(\ref{eq:deltacresult}) converges at $\omega \sim \min(\hbar/\tau_{\rm
D},T)$. We focus on the regime $T \gg \hbar/\tau_{\rm D}$, in which
the convergence is at $\omega \sim \hbar/\tau_{\rm D}$. In this regime
the inequality $\omega \ll T$ is obeyed for all frequencies $\omega$
contributing to the integral, so that all relevant 
interaction modes that contribute to dephasing can be described using 
the classical fluctuation-dissipation theorem. Indeed, one verifies
that in this regime the first-order interaction correction (\ref{eq:deltacresult})
agrees with the interaction correction to $\tilde c$ obtained in the
semiclassical framework of Sec.\ \ref{sec:motivation}, taken to first
order in the interaction propagator $L$.

Estimating the magnitude of the first-order correction
$\delta \tilde c_{ij}$ for $T \gg
\hbar/\tau_{\rm D}$, we find that $\delta \tilde c_{ij} \sim \tilde
c_{ij} T/T_{\phi}$, where $T_{\phi} \sim \hbar g/\tau_{\rm D}$ [see
Eq.\ (\ref{eq:Tphi}) above]. This
observation has two consequences: First, it implies that the regimes
of validity of first-order perturbation theory and the semiclassical
approach of Sec.\ \ref{sec:motivation} overlap: Both approaches are
valid if $\hbar/\tau_{\rm D} \ll T \ll T_{\phi}$. Second, it implies
that interactions give no significant correction to the weak
localization correction $\delta g^{\rm WL}$ if $T \lesssim
\hbar/\tau_{\rm D}$, so that we may ignore the difference between
the fully quantum-mechanical interaction correction $\delta \tilde
c_{ij}$ of Eq.\ (\ref{eq:deltacresult}) and the semiclassical result
in the low-temperature regime $T \lesssim \hbar/\tau_{\rm D}$ within
the limiting procedure outlined in Sec.\ \ref{sec:network}. (Both
approaches give essentially no interaction correction to weak
localization at these temperatures.) When combined, these two
observations justify the semiclassical considerations of Sec.\
\ref{sec:motivation}, as well as the expressions (\ref{eq:dGWL}) -- (\ref{eq:Gamma})
for the weak localization correction $\delta g^{\rm WL}$ that followed
from these considerations.

\begin{figure}
  \includegraphics{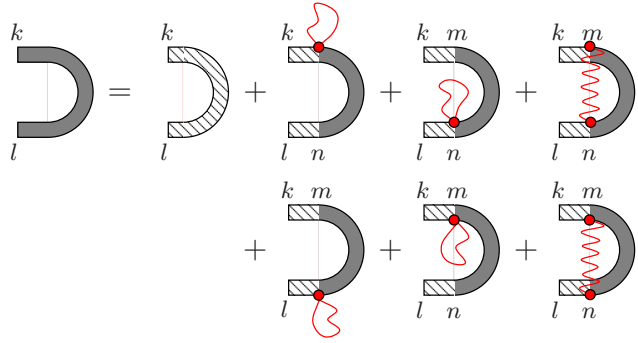}
  \caption{Dyson equation for the Cooperon obtained by perturbation 
  theory in the high temperature limit. The hatched boxes indicate 
  noninteracting Cooperon ladders, while gray shading indicates 
  that interactions are taken into account. Wiggly lines indicate 
  the equal time interaction propagator, which can either connect back  
  to the same propagation line, or to the opposite, time reversed one.
   \label{fg:dyson_diag}}
\end{figure}

For completeness, we mention that the full temperature dependence 
of $\delta g^{\rm WL} $ can also be obtained from 
diagrammatic perturbation theory. Following the above arguments, in
the limit $T \gg \hbar/\tau_{\rm D}$ all factors $\coth (\w / 2T) +
\tanh (\e - \w)/2T$ appearing in the calculation may be replaced by $2
T/\omega$, irrespective of the value of $\e$. This considerably
simplifies the calculation, and the $m$ interaction propagators that
appear in $m$th order in perturbation theory may then be placed 
independently of each other along the cooperon ladder.
Using Eq.\ (\ref{eq:LRapprox}) for the interaction propagator and 
writing the Cooperon ladders (without interaction corrections) in an 
integral form similar to Eq.\ (\ref{eq:cij0semi}), 
\begin{equation}
  \left( \tilde g + \tilde g_{\rm H}  + 2 \pi i \w \tilde \nu \right)^{-1} 
  \, = \, \left(  2 \pi \hbar \tilde \nu \right)^{-1} 
          \int_{0}^{\infty} d \tau  e^{- \tilde \gamma \tau - i \w \tau },
\end{equation}
one may perform the frequency integrations. The resulting expression consists
solely of time integrations with instantaneous interactions. The
remaining combinatorial problem leads to a Dyson equation of the form
shown in Fig.\ \ref{fg:dyson_diag}. Here the first term on the right
hand side is the noninteracting Cooperon 
$\tilde c_{kl} = (\tilde g + \tilde g_{\rm H})^{-1}_{kl}$
and the six other terms are obtained 
by different placements of the interaction propagators. 
[Note that where 
beginning and end are on the same Green's function line, an additional 
weight of $1/2$ arises from a factor $\int_{0}^{\infty} d \tau
\delta(\tau) = 1/2$.] 
Adding the six different contributions gives a vertex proportional to 
$ ( 4 \pi T / d_{\rm s} \hbar ) ( \tilde g^{-1}_{mm} + \tilde g^{-1}_{nn} 
- 2 \tilde g^{-1}_{mn} )$, so that one arrives at the Dyson equation
\begin{equation} 
  \tilde c_{kl} 
  \, = \, 
  \left( \tilde g + \tilde g_{\rm H} \right)^{-1}_{kl}
  - \sum_{m,n = 1 }^{\ND} \left[ \left( \Gamma + \Gamma_{\rm H} \right)^{-1} 
  \Gamma_{\phi}\right]_{km,ln} \tilde c_{mn},
  \label{eq:cijperturb}
\end{equation}
where $\Gamma$, $\Gamma_{\rm H}$, and $\Gamma_{\phi}$ are rank-four
tensors whose definition is given below
Eq.\ (\ref{eq:Gamma}). 
With a little algebra, one verifies that Eq.\ (\ref{eq:cijperturb}) is
equivalent to the result (\ref{eq:cijdef}) derived using semiclassical
arguments.

Equation (\ref{eq:deltacresult}) can also be used to calculate the
magnitude of energy quanta $\omega$ exchanged with the fluctuating
electromagnetic field in the quantum dots. Hereto, we note that the
sum of the second and fourth terms between brackets  $\{ \ldots \}$ 
in Eq.\ (\ref{eq:deltacresult}) 
is proportional to (minus) the probability $p_1(\omega)$
for emission or absorption of a photon along the electron's 
trajectory, so that
\begin{eqnarray}
p_1(\w) & = & 
  \frac{1}{g'_{1 \cdot} \tilde g^{-1}_{\cdot \cdot}  g'_{\cdot 2}}
  \sum_{\alpha, \beta = 1}^{\ND} \frac{ \w }{ 2 \pi T \sinh^2{\w / 2 T }}
  \Ima[ 4 \pi^2 \nu_{\alpha } \nu_{\beta} L^{\rm R}_{\alpha \beta}(\w) ]\,
  \Rea \big[ g'_{1 \cdot} 
     \tilde g^{-1}_{\cdot \alpha} 
   \left( \tilde g + i 2 \pi \w \tilde \nu \right)^{-1}_{\alpha \beta}
   \tilde g^{-1}_{\beta \cdot} g'_{\cdot 2}  \big]
  \nonumber \\ &=&
  \frac{16 T \pi^2}{g'_{1 \cdot} \tilde g^{-1}_{\cdot \cdot}
   g'_{\cdot 2}}
   \sum_{\alpha, \beta = 1}^{\ND} 
    \nu_{\alpha} \tilde g^{-1}_{\alpha \beta} \nu_{\beta} \,
  \mbox{Re}\, [ g'_{1 \cdot} 
     \tilde g^{-1}_{\cdot \alpha} 
   \left( \tilde g + i 2 \pi \w \tilde \nu \right)^{-1}_{\alpha \beta}
   \tilde g^{-1}_{\beta \cdot} g'_{\cdot 2}],
  \label{eq:probemitone}
\end{eqnarray}
where, in the second equality, we took the limit
$T \gg \hbar/\tau_{\rm D}$. 
The probability that one inelastic scattering event 
of arbitrary frequency occurs is $P_1 = \int d \w p_1(\w)$.
Equation (\ref{eq:probemitone}) is valid as long as $P_1 \ll 1$,
 so that first-order perturbation theory is
sufficient.

From Eq.\ (\ref{eq:probemitone}) we conclude that the energy of
photons that are emitted or absorbed is limited by
$\min(\hbar/\tau_{\rm D}, T)$. The temperature $T_{\phi}$ at which the
interaction correction to weak localization becomes relevant is the
temperature at which the probability that at least one energy quantum
is exchanged becomes of order unity. However, the typical exchanged
energy remains of order $\hbar/\tau_{\rm D}$ for all temperatures.
This implies that the broadening of the distribution function by
inelastic processes is parametrically smaller than the temperature
$T$, by a factor $1/g \ll 1$. Transport in the quantum dot network is
thus quasielastic for all temperatures. (Inelastic processes become
relevant only if $T \gtrsim E_{{\rm Th},i} g^{1/2}$, where $E_{{\rm
    Th},i}$ is the Thouless energy of an individual quantum dot.)

\subsubsection{Interaction corrections to the conductance}

The relevant diagrams for the interaction correction to the
conductance $\delta g^{\rm int}$ are shown in Fig.\
\ref{fg:AA_box_arg}. These diagrams do not involve Cooperon propagators.
The diagram shown in Fig.\ \ref{fg:AA_box_arg}a is analogous 
to the ones
we have already encountered in calculating the (first-order) dephasing
correction to weak localization. 
It gives an interaction correction to the diffuson
propagator $\tD(\e,\e)$ that depends on the frequency $\omega$ of
the interaction propagator,
\begin{eqnarray}
  \delta \tD_{\beta \alpha, ij}(\omega)^{(a)} &  = &
  \tD_{i\cdot}(\e,\e) I^{\rm R}_{\beta,\cdot\cdot}
  \tD_{\cdot\cdot}(\e-\w,\e) I^{\rm R}_{\alpha,\cdot\cdot}
  \tD_{\cdot j}
  \nonumber \\ 
  & = & -  \frac{ 4 M_i \nu_\be}{\nu_i}   
  \tilde{g}^{-1}_{i \be} 
  \left( \tilde{g} + i 2 \pi \w \tilde{\nu} \right)^{-1}_{\be \al} 
  \tilde{g}^{-1}_{\al j}
  \frac{ 4 M_j \nu_{\al}}{\nu_j}
\end{eqnarray}
(The frequency $\w$ will be integrated over in the final expression.)
For the remaining diagrams, we need to consider an interaction vertex
that connects an advanced and a retarded Green function. Such an
interaction vertex is dressed by a Diffuson propagator, which allows
the interaction vertex to be placed in a dot different from the one
that appears at the outer end of the dressed interaction
vertex,
\begin{eqnarray}
  \tilde I_{\alpha,i}(\omega) &=&
  \delta_{\alpha i}
  + \sum_{k} D_{ik}(\e-\w,\e)
  \mbox{tr}\, \langle \GA_{k\alpha}(\e) \rangle
  \langle \GR_{\alpha k}(\e-\w) \rangle
  \nonumber \\ &=&
  \frac{4 M_i \nu_{\alpha}}{\nu_i}
  \left(\tilde g +  i 2 \pi \w \tilde \nu \right)^{-1}_{i \alpha}.
\end{eqnarray}
\begin{figure}
  \includegraphics{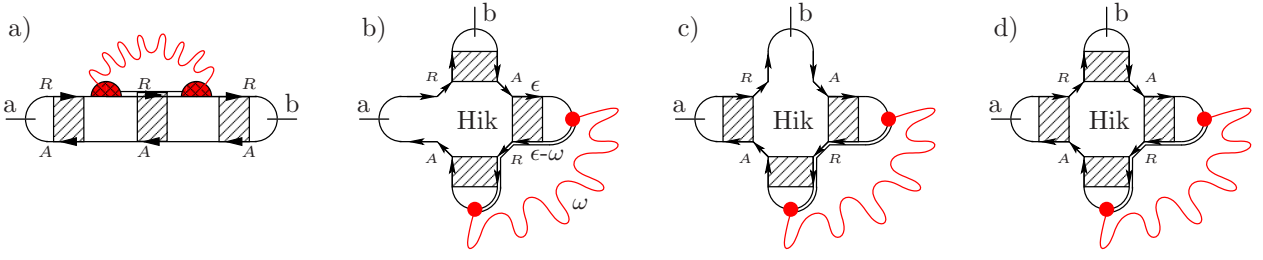}
  \caption{Diagrams contributing to $\delta g^{\rm int}$. 
  The Hikami box is defined in Fig.\ \ref{fg:Gni_ext}.
\label{fg:AA_box_arg}}
\end{figure}%
With this interaction vertex, the diagrams of Fig.\
\ref{fg:AA_box_arg}b--d (without the outer diffusion ladders) can be 
represented by Hikami boxes $B_{ij,kl}(\w)$ and $B'_{aj,kl}$ of
Eqs.\ (\ref{eq:Hikami}) and (\ref{eq:HikamiL}), but with $g_{\rm H}
\to 0$ because no Cooperon ladders are involved. 
Combining the two interaction contributions to the interaction
correction we find 
\begin{eqnarray}
  \delta g^{\rm int} & = 
  & - 4 \, \int \frac{ d \w }{2 \pi}  
   \left( \frac{ \partial }{ \partial \w }  \w \coth \frac{ \w }{ 2 T} \right)
  \sum_{\ab}^{N_{\rm D}} \sum_{k,l = 1}^{N_{\rm D}}
  \Ima \Big[ \LA_{\ab}(\w)  
  D_{1 \cdot}'  \delta \tD_{\beta \alpha, \cdot \cdot}(\omega)^{(a)}
  D_{\cdot 2}'
  \nonumber \\ && \mbox{}
  + \LA_{\ab}(\w) \tilde I_{k\alpha} \tilde I_{l\beta}
  \left( B_{1l, \cdot k}' \tD_{\cdot\cdot} D_{\cdot 2}' +
  D_{1 \cdot}' \tD_{\cdot\cdot} B'_{\cdot l,2k} +
  D_{1 \cdot}' \tD_{\cdot\cdot} B_{\cdot l,\cdot k}(\w)
  \tD_{\cdot\cdot} D_{\cdot 2}' \right)
  \Big].
\end{eqnarray}
\end{widetext}
Expressing the propagators in terms of the matrices $\tilde g$ and
$\tilde f$, we find that $\delta g^{\rm int}$ naturally separates
into two contributions, which are given by equations
(\ref{eq:AAgen1})--(\ref{eq:AAgen2c}) of Sec.\
\ref{sec:network}. Both corrections are small for all temperatures,
and it is not necessary to consider higher order contributions 
involving more than one interaction propagator $L$.

\section{Application to double quantum dot}
\label{sec:double}

We now apply the theory of the previous sections to the case of a double
quantum dot. There are two cases of interest: A linear configuration,
in which each dot is coupled to one reservoir, see Fig.\
\ref{fg:dd_circ}a, and a side-coupled configuration, in which
both reservoirs are connected to the same quantum dot, see Fig.\
\ref{fg:dd_circ}b.

\subsection{Linear configuration}

\begin{figure}
\includegraphics{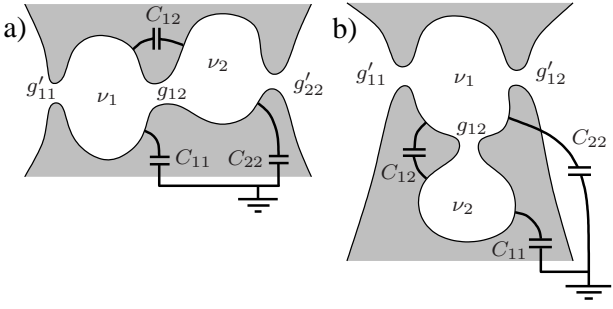}
\caption{Schematic drawings of two double quantum dots. 
Panel (a) shows a linear
  configuration; Panel (b) shows a side-coupled
  configuration.  \label{fg:dd_circ}}
\end{figure}

The conductance matrix for the linear double quantum dot reads
\begin{eqnarray}
  \tilde{g} & = & 
  \left( \begin{array}{cc}  \go + \gc  &  - \gc \\ - \gc & \gt + \gc  \end{array} \right) \mbox{,}
\end{eqnarray}
where $g_{11}'$ and $g_{22}'$ are the dimensionless conductances of the
contacts connecting the two dots to the reservoirs, and $\gc$ is
the dimensionless conductance of the contact between the two dots, see
Fig.\ \ref{fg:dd_circ}. The form factor matrix $\tilde{f}$ has a
similar structure, with $\go$, $\gt$, and $\gc$ replaced by
$\fo$, $\ft$, and $\fc$, respectively.
The classical conductance of the system is $\Gcl = (d_{\rm s}  e^2/h) \gcl$, with
\begin{eqnarray}
  \gcl^{-1} = \go^{-1} + \gt^{-1} + \gc^{-1},
\end{eqnarray}
see Sec.\ \ref{sec:network}, Eq.\ (\ref{eq:gcl}). 

\subsubsection{Weak localization}

The zero temperature
weak localization correction to the conductance $\delta G^{\rm WL}
= (d_{\rm s} e^2/h) \delta g^{\rm WL}$ follows from substitution of the
zero-temperature Cooperon $\tilde c(0)$ of Eq.\
(\ref{eq:codlimit}) into Eq.\ (\ref{eq:dGWL}),
\begin{eqnarray} 
  \frac{\delta g^{\rm WL}}{\gcl^2} &=&
  -
  \frac{\fo/\go^2 + \fc/\gc^2}{\go + \gho + \gc 
  - \gc^2/( \gt + \ght + \gc)}
  \nonumber \\ && \mbox{}
  -
  \frac{\ft/\gt^2 + \fc/\gc^2}{\gt + \ght + \gc 
  - \gc^2/( \go + \gho + \gc)}
  \nonumber \\ && \mbox{}
  - \frac{2 (\gc - \fc)/\gc}
  { ( \go + \gho + \gc ) ( \gt + \ght + \gc) - \gc^2 }.
  \nonumber \\
  \label{eq:ddwl}
\end{eqnarray}
Here $\ght$ and $\gho$ are dimensionless numbers describing the effect
of an applied magnetic field, see Eq.\ (\ref{eq:gH}). The limit of
zero magnetic field $\ght = \gho = 0$ agrees with the result obtained
previously by Golubev and Zaikin.\cite{golubev2006} 
The high-temperature limit of
$\delta g^{\rm WL,d}$ of the weak localization correction 
is found by taking the diagonal
contribution $\tilde c^{\rm d}$ of Eq.\ (\ref{eq:cdlimit}) for the 
Cooperon propagator,
\begin{eqnarray}
  \frac{\delta g^{\rm WL,d}}{\gcl^2} &=&
  - \frac{ \fo/  \go^2  +  \fc /\gc^2 }{ \go + \gho + \gc }
   - \frac{ \fc/\gc^2 +  \ft/\gt^2}{ \gt+ \ght + \gc }
  . \label{eq:sdwl}
  \nonumber \\
\end{eqnarray}
Note that $|\delta g_{\rm WL}^{\rm d}| < |\delta g_{\rm WL}|$. 
The remainder of the weak localization correction, $\delta g^{\rm
WL} - \delta g^{\rm WL,d}$, is temperature dependent because
of dephasing from electron-electron interactions.
Taking the temperature-dependent Cooperon from Eq.\
(\ref{eq:cijdef}), we find that the temperature dependence of the
full matrix $\tilde c(T)$ is encoded in a single scalar function $f(T)$,
\begin{equation}
  \tilde c(T) = \tilde c(0) - [\tilde c(0) - \tilde c^{\rm d}] f(T).
  \label{eq:ddcdeph}
\end{equation}
Equation (\ref{eq:ddcdeph}) immediately implies that
\begin{equation}
  \delta g^{\rm WL}(T) = \delta g^{\rm WL,d}
  + [\delta g^{\rm WL}(0) - \delta g^{\rm WL,d}][1 - f(T)],
  \label{eq:ddSplitTemp}
\end{equation}
where $\delta g^{\rm WL}(0)$ and $\delta g^{\rm WL,d}$ are given
in Eqs.\ (\ref{eq:ddwl}) and (\ref{eq:sdwl}), respectively.
In the regime where temperature is large enough for dephasing effects
to give a sizeable correction to the weak localization correction to the
conductance, we obtain $f(T)$ from 
Eq.\ (\ref{eq:cijdef}),
\begin{equation}
  f(T) = \frac{ T }{ T_{\phi} + T},
  \label{eq:fTdys}
\end{equation}
with
\begin{equation}
  \frac{ T_{\phi}}{ d_{\rm s}}  = 
   \frac{\hbar (\tauo + \taut)(\go \gt + \go \gc + \gt \gc)}{4 \pi \tau_+ \tau_- ( \go + \gt)}.
  \label{eq:Tzsc}
\end{equation}
Here $\tauo$ and $\taut$ are the (classical) dwell times of
the two dots, modified for the presence of a magnetic field,
\begin{eqnarray}
  \tauo & = & \frac{ 2 \pi \hbar \nu_1}{ \go + \gho + \gc}  \mbox{,  } \, 
  \taut \, = \,  \frac{ 2 \pi \hbar \nu_2}{ \gt + \ght + \gc},
\end{eqnarray}
whereas $\tau_{\pm}$ are time scales representing the relaxation of
symmetric ($+$) or antisymmetric ($-$) charge configurations in the
double dot,
\begin{equation}
  \frac{ 1}{ \tau_{\pm} }  =  \frac{1}{2 \tauo}  + \frac{ 1 }{ 2\taut } 
  \mp \frac{1}{2} \sqrt{ \left( \frac{ 1}{\tauo} - \frac{ 1}{\taut}
  \right)^2 + \frac{ \gc^2}{ \pi^2 \hbar^2 \nu_1 \nu_2} } .
  \label{eq:taupm}
\end{equation}

It is instructive to compare Eq.\ (\ref{eq:fTdys}) with the expression 
for $f(T)$ obtained in first-order perturbation theory,
\begin{eqnarray}
  f(T)  & = & \int  \frac{ d \w}{2 \pi}  \frac{ \w / 2 T}{\sinh^2( \w/2T)} \frac{2 \w^2}{ ( 1 + \w^2 \tau_+^2/\hbar^2  ) ( 1 + \w^2 \tau_-^2/\hbar^2 ) } \, \nonumber \\
&& \mbox{} \times \frac{ \tau_+^3 \tau_-^3}{\hbar^4 \tau_1 \tau_2} 
  \mbox{Im}\, [L^{\rm R}_{11}(\w) + L^{\rm R}_{22}(\w) - 2 L^{\rm
      R}_{12}(\w)].
  \nonumber \\
\label{eq:fT}
\end{eqnarray}
The integral in Eq.\ (\ref{eq:fT}) converges for frequencies $\omega/\hbar$
of order $\tau_{\pm}^{-1}$. For these frequencies, we may neglect the
capacitance $C$ in the expression for the interaction propagator $L$
since $C/e^2 \ll \nu$. 
The resulting frequency integration yields
\begin{eqnarray}
  f(T) & =
   &  \frac{  2 \pi T \tp \tm }{ 3 \hbar ( \tp - \tm ) }\frac{T }{ T_{\phi}}
   \nonumber \\ && \mbox{} \times
  \left[ {\cal F}_1( 2 \pi T \tau_- /\hbar) - {\cal F}_1( 2 \pi T \tau_+ /\hbar)\right]  \label{eq:fTlindd}
\end{eqnarray}
where
\begin{equation}
  {\cal F}_1(x) =  \frac{3}{ x^2} \left\{ 
  \frac{ 1}{ x} \left[ 2 \psi'\left(\frac{1}{x}\right) - x^2 \right] 
  - 2  \right\},
  \label{eq:F1}
\end{equation}
$\psi'$ being the derivative of the digamma function. 
With the asymptotic behavior of ${\cal F}_1(x)$,
\begin{equation}
  {\cal F}_1(x) = \left\{ \begin{array}{ll} 
  1 - \frac{1}{5} x^2 + \frac{1}{7} x^4 + \ldots, & \, x \ll 1, \\
  \frac{3}{x} - \frac{ 6}{ x^2} + \frac{ \pi^2}{x^3} + \ldots,  & \, x \gg 1 ,
\end{array} \right.
\end{equation}
we identify three different regimes for the temperature dependence of
the dephasing correction:
\begin{equation}
  f(T) = 
  \frac{ 1}{ 15} \tp \tm ( \tp + \tm ) \left( \frac{ 2 \pi T }{ \hbar}\right)^3  \frac{ T }{ T_{\phi}} \label{eq:fTlow}
\end{equation}
if $T \ll \hbar /\tau_{+}$,
\begin{equation}
  f(T)  = 
   \frac{ 2 \pi T  \tau_- }{3 \hbar} \frac{ T}{ T_{\phi} }\label{eq:fTmed}
\end{equation}
if $\hbar / \tau_{+} \ll T \ll \hbar / \tau_{-}$, and
\begin{eqnarray}
  f(T) = T/T_{\phi}
  \label{eq:fThigh}
\end{eqnarray}
if $ \hbar / \tau_{-} \ll T $, where $T_{\phi}$ 
is given by Eq.\ (\ref{eq:Tzsc})  above. 
The intermediate temperature regime exists
only if $\tau_{+} \gg \tau_{-}$. A comparison of Eq.\
(\ref{eq:fThigh}) with Eqs.\ (\ref{eq:fTdys}) shows that the two
expressions for $f(T)$ agree in the temperature regime $\hbar/\tau_-
\ll T \ll T_{\phi}$ where both expressions are valid. It is in  this
temperature regime that the factor $(\omega/2 T)/\sinh^2(\omega/2T)$
in Eq.\ (\ref{eq:fT}) can be approximated by $2 T/\omega$,
which is the appropriate weight appearing in the classical
fluctuation-dissipation theorem.

It should be noted that the low temperature corrections, 
Eqs.\ (\ref{eq:fTlow}) and (\ref{eq:fTmed}), result in contributions
to the conductance of order $\mathcal{O}(1/g)$. Such contributions are
beyond the accuracy achieved in the limiting procedure outlined in
Sec.\ \ref{sec:network}. Further contributions 
of the same order might be obtained by calculating, {\em e.g.}, 
weak localization corrections to the interaction corrections $\delta
g^{{\rm int},1}$ and $\delta g^{{\rm int},2}$. For disordered metals 
such contributions have been considered explicitly in Ref.\ 
\onlinecite{aleiner1999}. 

The above equations take a simpler form in the limiting cases of 
large and small interdot coupling $\gc$ and of a large magnetic
field. For small interdot coupling
$\gc \ll \min(\go,\gt)$, one has 
\begin{eqnarray}
  \delta g^{\rm WL} & = & 
   - \frac{\fo \gc^2 + \fc \go^2}{\go^2(\go
  + \gho)}  -  \frac{\ft \gc^2 + \fc \gt^2}{\gt^2(\gt + \ght)} 
  \nonumber \\ && \mbox{} 
  - \frac{2(\gc - \fc)\gc}{(\go + \gho)(\gt + \ght)} 
  \frac{T_{\phi}}{T_{\phi} + T}, ~~~~ \\
 \frac{  T_{\phi}}{ d_{\rm s} }  &=& \go\gt\frac{(\go + \gho)\nu_1^{-1} + (\gt + \ght)\nu_2^{-1}}
  {8 \pi^2 (\go + \gt)}, ~~~~
\end{eqnarray}
so that only a small part of the total weak localization correction is
temperature dependent. In the limit of a large interdot conductance,
$\gc \gg \max(\go,\gt,\gho,\ght)$, the full weak localization
correction acquires a temperature dependence,
\begin{eqnarray}
  \delta g^{\rm WL }  & = &
  -  \frac{ \gt^2 \fo + \go^2 \ft}{( \go + \gt)^2( \go  + \gho + \gt +
  \ght) } \nonumber \\ && \mbox{} \times \frac{T_{\phi}}{T_{\phi} +
  T},
  \nonumber  \\
  \frac{ T_{\phi}}{ d_{\rm s} }  &=& \gc
  \frac{(\go + \gho + \gt + \ght)(\nu_1^{-1} + \nu_2^{-1})}{8 \pi^2}.
  ~~~~
  \label{eq:dgWLstrong}
\end{eqnarray}
Finally, in the limit of large magnetic field, $\gho,\ght \gg
\max(\go,\gt,\gc)$, we have
\begin{eqnarray}
  \delta g^{\rm WL} & = & - \gcl^{2}
  \frac{\fo/\go^2 + \fc/\gc^2}{\gho} 
  - \gcl^{2} \frac{\ft/\gt^2 + \fc/\gc^2}{\ght} 
  \nonumber \\ && \mbox{}
  - \gcl^2 \frac{ \gc - \fc}{\gc \gho \ght} \frac{T_{\phi}}{T_{\phi} +
  T}, \\
  \frac{ T_{\phi}}{ d_{\rm s}}  &=& \frac{\gc}{ 8 \pi^2 } 
  \left( \gho \nu_1^{-1} + \ght \nu_2^{-1} \right).
\end{eqnarray}

A special case of two weakly coupled quantum dots ($\gc \ll \go,\gt$)
with tunneling contacts ($\fo \ll \go$, $\ft \ll \gt$, $\fc \ll \gc$)
has been considered recently
by Golubev and Zaikin.\cite{golubev2007} 
While our calculation agrees with that of Ref.\
\onlinecite{golubev2007} in the high temperature regime $T \gg T_{\phi}$,
significant differences appear in the low temperature limit. In
particular, Golubev and Zaikin find a finite dephasing correction to
weak localization at zero temperature, whereas we find no such
effect.
A similar discrepancy has been found previously in the context of 
dephasing from the electron-electron interaction in 
disordered metals.\cite{golubev1998,aleiner1999} In this case 
the neglect of recoil effects in the influence functional approach
used by Golubev and Zaikin has been identified as the cause of the
problem.\cite{vondelft2005} This causes an ultraviolet divergence,
which does not appear in the perturbation theory, where it is avoided
by the $\tanh$-term in the factor $\coth({ \w/2T}) + \tanh{((\e -
  \w)/2T)}$ that sets the magnitude of the dephasing correction at low
temperatures, see, {\em e.g.}, Eq.\ (\ref{eq:Gdeph}) and Refs.\
\onlinecite{aleiner1999, vondelft2005}. (Neglect of recoil amounts to
neglecting the $\w$-dependence of the argument of the $\tanh$, which
causes this factor to no longer 
approach zero at large frequencies $\w$.)
We believe that the discrepancy between our result and that 
of Ref.\ \onlinecite{golubev2007} has the same origin.

\subsubsection{Interaction corrections}

The interaction corrections $\delta g^{{\rm int},1}$ and $\delta
g^{{\rm int},2}$ do not depend on the magnetic field. Hence, the 
relevant time scales do not involve $\gho$ and $\ght$, and we define
\begin{eqnarray}
  \tau_{i} = \frac{2 \pi \hbar \nu_i}{g_{ii}' + \gc},\ \ i=1,2.
\end{eqnarray}
Again, we introduce time scales $\tau_{\pm}$ related to $\tau_{1}$ and
$\tau_{2}$ as in Eq.\ (\ref{eq:taupm}) above.
For the first interaction correction
$\delta g^{{\rm int},1}$ we then find
\begin{eqnarray}
  \delta g^{{\rm int},1} &=&  
  \frac{ \gcl^3}{ d_{\rm s} \go \gc \gt}  
  \int d \w  
   \left( \frac{\partial }{ \partial \w } \w \coth  \frac{ \w}{2T} \right)
  \nonumber \\ & & 
  \times  \Ima \frac{(\tp + \tm )/\hbar }{ (1 + i \w \tp/\hbar)( 1 + i \w \tm /\hbar)}.
  \label{eq:dg1int}
\end{eqnarray}
This result was obtained previously in Ref.\ \onlinecite{brouwer2008a}
for the symmetric case $\go = \gt$, $\nu_1 = \nu_2$ and in Ref.\
\onlinecite{golubev2004b} for the case $\go = \gt = \gc$, $\nu_1 = \nu_2$.
The frequency integral in Eq.\ (\ref{eq:dg1int}) can be 
evaluated in terms of digamma functions. We have
\begin{widetext}
\begin{eqnarray}
&&  \int  d \w \left( \frac{\partial }{ \partial \w } 
   \w \coth  \frac{ \w}{2T} \right)
   \Ima 
  \left[ \frac{1}{ (1 + i \w \tau_{\alpha} / \hbar)( 1 + i \w
  \tau_{\beta} /\hbar)} \right]
  =
  \frac{ 2 \hbar }{  \tau_{\alpha } - \tau_{\beta} } 
   \left[ {\cal F}_2\left(\frac{\hbar}{ 2 \pi T \tau_{\alpha} }\right) 
    - {\cal F}_2\left( \frac{\hbar }{2 \pi T \tau_{\beta}} \right)
   \right],
  ~~~~~~~~
  \label{eq:F2integ}
\end{eqnarray}
where
\begin{equation} 
  {\cal F}_2(x) \, = \, \psi(1 + x) + x \psi'(1 + x)
\end{equation}
and $\psi(x)$ is the digamma function.\cite{golubev2004b} From the asymptotic behavior
of ${\cal F}_2$, 
\begin{equation}
  {\cal F}_2(x) = \left\{ \begin{array}{ll} 
  - \gamma  + \frac{\pi^2 }{3} x - 3 \zeta(3) x^2 + \ldots, & \, x \ll 1, \\
  1 + \ln x + \frac{1}{12 x^2} + \ldots,  & \, x \gg 1 ,
\end{array} \right.
\end{equation}
with $\gamma$ the Euler-Mascheroni constant, we obtain the high and low temperature limit of the interaction 
correction $\delta g^{{\rm int},1} $
\begin{equation}
  \delta g^{{\rm int},1}  = 
  - \frac{2 \gcl^3}{  d_{\rm s} \go \gc \gt} \times
  \left\{ \begin{array}{ll}
  \displaystyle{
   \frac{ \tp + \tm }{ \tp - \tm  \vphantom{M^M_M}}
  \ln \frac{\tp}{\tm}}, & T \ll \hbar/\tau_{\pm}, \\
  \displaystyle{
  \frac{\pi \hbar (\tp + \tm) \vphantom{M^M_M}}{6 T\tp\tm}}, &
  T \gg \hbar/\tau_{\pm}.
  \end{array} \right.
\end{equation}

The second interaction correction $\delta g^{{\rm int},2}$ is
expressed in terms of interaction-induced shifts $\delta \go$, $\delta
\gt$, and $\delta \gc$ to the conductances $\go$, $\gt$, and $\gc$,
respectively, see Eq.\ (\ref{eq:AAgen2}). In contrast to the
interaction correction $\delta g^{{\rm int},1}$ considered above,
the frequency integrations needed to calculate $\delta \go$, $\delta
\gt$, and $\delta \gc$ converge only if we account for the finite 
(nonzero) capacitances of the quantum dots, see Eq.\
(\ref{eq:AAgen2c}). [The integration in Eq.\ (\ref{eq:AAgen2c})
diverges logarithmically if the limit $C_{ii}/e^2 \nu_{i} \to 0$ is taken.]

Below we give 
explicit expressions for the case of a symmetric double dot only, 
$\go=\gt=g'$, $\fo=\ft=f'$, 
$\nu_1 = \nu_2$, and $C = C_{11} = C_{22}$. In this case, the
logarithmic divergence of the integration in Eq.\ (\ref{eq:AAgen2a})
is cut off at the inverse of the charge-relaxation times,
\begin{eqnarray}
\tau_{\rm c +  } & = 
& \frac{ \tp}{d_{\rm s} e^2 \nu / C}, \ \
\tau_{\rm c - } 
 = \frac{ \tm }{d_{\rm s} e^2 \nu / ( C + 2 C_{12} ) } \mbox{,} ~~~~~
\end{eqnarray}
and the corrections $\delta \go = \delta \gt = \delta g'$ and $\delta
\gc$ are found to be
\begin{eqnarray}
\delta g'  & = &
   \frac{ g' - f' }{ d_{\rm s} g' } \sum_{\sigma=\pm} \frac{ \tau_{\sigma}}{\tp}
   \left[ {\cal F}_2 \left(\frac{\hbar}{ 2 \pi T \tau_{\sigma}  }\right)
   - {\cal F}_2\left(\frac{\hbar}{ 2 \pi T \tau_{{\rm c} \sigma} }\right) \right] ,
  \label{eq:dgcontact}
 \\
\delta \gc  & = &
    \frac{2 (  \gc - \fc)}{ d_{\rm s} \gc} \frac{ \tp - \tm }{ \tp} 
\left[ {\cal F}_2\left( \frac{ 1}{ 2 \pi T \tm / \hbar} \right) - 
  {\cal F}_2 \left(\frac{1}{2 \pi T \tau_{{\rm c}-} / \hbar} \right) \right]
   \mbox{.}
  \label{eq:dgmiddle}
\end{eqnarray}
For the case $g' = g_{12}$, $f' = f_{12}$ and $ C_{12}= 0$, Eqs.\ (\ref{eq:dgcontact}) and
(\ref{eq:dgmiddle}) agree with results obtained previously in 
Ref.\ \onlinecite{golubev2004b}.
[ The result of Ref.\ \onlinecite{golubev2004b} differs from 
Eqs.\ (\ref{eq:dgcontact}) and (\ref{eq:dgmiddle}) if $C_{12} > 0$ because 
Ref.\ \onlinecite{golubev2004b} includes cross capacitances between
 each dot and adjacent reservoir of the same magnitude as the 
cross capacitance $C_{12} $ between the two dots.]
Equation (\ref{eq:dgcontact}) simplifies to the renormalization of the
contact conductance for a single quantum dot 
in the limit $\gc \to \infty$.\cite{golubev2004a,brouwer2005,brouwer2005d}
Again making use of the asymptotic behavior of the digamma function,
we find that the above expressions simplify to
\begin{equation}
  \delta g'  = 
  - \frac{g' - f'}{ d_{\rm s} g'}\times
  \left\{ \begin{array}{ll}
  \displaystyle{
     \ln \frac{ \tp }{ \tau_{{\rm c} +}} 
   + \frac{ \tm }{ \tp} \ln \frac{ \tm }{ \tau_{{\rm c} -}} }
  , & T \ll \hbar/\tau_{\pm}, 
  \\
  \displaystyle{ 
    \ln \frac{ e^{1 + \gamma}}{2 \pi T \tau_{{\rm c} +}}  
  + \frac{ \tm}{\tp} \ln \frac{ e^{1 + \gamma}}{2 \pi T \tau_{{\rm c} -}} }
  , &   \hbar / \tau_{\pm } \ll T \ll \hbar/\tau_{{\rm c}\pm}, 
  \\ 
  \displaystyle{
  \frac{\pi \hbar}{6 T \tp} 
  \left( \frac{ \tp }{ \tau_{{\rm c} +}} + \frac{\tm}{ \tau_{{\rm c} - }} \right) } , &
   \hbar/\tau_{{\rm c}\pm} \ll T, 
  \end{array} \right.
\end{equation}

\begin{equation}
  \delta \gc  = 
  - \frac{4 ( \gc - \fc)}{ d_{\rm s} \gc} \frac{ \tp - \tm }{ \tp} \times
  \left\{ \begin{array}{ll}
  \displaystyle{
   \ln \frac{\tm}{ \tau_{{\rm c} -}} }, & T \ll \hbar/\tau_{\pm}, 
  \\
  \displaystyle{ 
    \ln \frac{ e^{1 + \gamma}}{2 \pi T \tau_{{\rm c} -}} }
  , &   \hbar / \tau_{-}  \ll T \ll \hbar/\tau_{{\rm c}-}, 
  \\
  \displaystyle{
  \frac{\pi \hbar }{6 T\tau_{{\rm c}-}} }, &
   \hbar/\tau_{{\rm c}\pm} \ll T .
  \end{array} \right.
\end{equation}

\subsection{Side-coupled quantum dot}

For the side-coupled double dot configuration of figure
\ref{fg:dd_circ} the structure of the weak localization correction
and the interaction corrections is essentially the same as for the
linear configurations. 
The classical conductance is 
\begin{equation}
  g_{\rm cl}^{-1} = \go^{-1} + \gts^{-1}.
\end{equation}
The weak localization correction to the conductance is
\begin{eqnarray}
  \delta g^{\rm WL} &=&
  - \frac{\ft \go^2 + \fo \gts^2}{(\go + \gts)^2(\go + \gts + \gc +
  \gho)}
  \left\{ 1 + \frac{\gc^2 [1 - f(T)]}{(\go + \gts + \gho)(\gc + \ght) +
  \gc \ght} \right\}, \nonumber
\end{eqnarray}
\end{widetext}
where $f(T) = T/(T_{\phi} + T)$,
\begin{equation}
  \frac{ T_{\phi}}{ d_{\rm s}}  \, = \, \frac{ 1 }{ 4 \pi } \frac{\tauo + \taut}{ \tau_+ \tau_-}  
   \gc,
  \label{eq:Tzscmain}
\end{equation}
and
\begin{eqnarray}
  \tauo & = & \frac{ 2 \pi \hbar \nu_1}{ \go + \gts + \gho + \gc}  \mbox{,  } \, 
  \taut \, = \,  \frac{ 2 \pi \hbar \nu_2}{ \gc + \ght} , ~~~ \label{eq:tausc}
\end{eqnarray}
with $\tau_{\pm}$ given in terms of $\tauo$ and $\taut$ as in
Eq.\ (\ref{eq:taupm}). 

Again, it is instructive to compare to what one finds to lowest
order in perturbation theory. The result is identical to Eq.\ ({\ref{eq:fTlindd}}), 
where $\tau_1, \tau_2$ and $T_{\phi}$ are those of the 
side-coupled system, Eqs.\ (\ref{eq:Tzscmain}) and (\ref{eq:tausc}). 
Simplified expressions for the function $f(T)$
in the regimes
 $T \ll \hbar/\tau_+$, $\hbar/\tau_+ \ll T \ll \hbar/\tau_-$, and
$\hbar/\tau_- \ll T $ 
are as in Eqs.\
(\ref{eq:fTlow})--(\ref{eq:fThigh}). 

In the limit of small interdot coupling $\gc \to 0$ only a very small
fraction of the weak localization correction is temperature dependent,
\begin{eqnarray}
   \delta g^{\rm WL} &=&
  - \frac{\ft \go^2 + \fo \gts^2}{(\go + \gts)^2(\go + \gts + \gho)}
  \nonumber \\ && \mbox{} \times
  \left[1 + \frac{\gc^2}{(\go + \gts + \gho) \ght} 
  \frac{T_{\phi}}{T_{\phi} + T}\right], \nonumber \\
  \frac{ T_{\phi}}{d_{\rm s}}  &=& \frac{\gc}{ 8 \pi^2 } 
  \left[ (\go + \gts + \gho)\nu_1^{-1} + \ght \nu_2^{-1} \right].
  ~~~~~
\end{eqnarray}
In the opposite limit of a large interdot conductance the
entire weak localization correction is temperature dependent. In this
limit there is no difference between the linear and side-coupled
configurations, and one finds that $\delta g^{\rm WL}$ is given by
Eq.\ (\ref{eq:dgWLstrong}) above, with $\gt$ replaced by $\gts$.
Finally, in the limit of large magnetic fields we find
\begin{eqnarray}
\delta g_{\rm WL} &=& - \frac{ \gts^2 \fo + \go^2 \ft}{(\go + \gts)^2
  \gho} 
  \left(1 + \frac{\gc^2}{\gho\ght}\frac{T_{\phi}}{T_{\phi} + T} \right),
  \nonumber \\
  \frac{T_{\phi}}{ d_{\rm s}} & = &  \frac{\gc}{ 8 \pi^2} 
  \left( \gho \nu_1^{-1} + \ght \nu_2^{-1} \right).
\end{eqnarray}

With a side coupled quantum dot, the interaction correction 
$\delta g^{{\rm int},1}$ to the conductance vanishes. The interaction
correction $\delta g^{{\rm int},2}$ coming from the renormalization
 of the contact conductances remains. 
The detailed expressions are rather lengthy and will not be
reported here. 

\section{Conclusion}
\label{sec:conc}

We have calculated the quantum corrections to the conductance of a
network of quantum dots, including the full dependence on temperature
and magnetic field. Our results are valid in the limit that the
quantum dot network has conductance $g$ much larger than the conductance
quantum, so that the quantum corrections are small in comparison to
the classical conductance, and in the limit that the electron dynamics
inside each quantum dot is ergodic. Following the literature, we
separated the quantum corrections into the weak localization correction
$\delta g^{\rm WL}$ and two interaction corrections $\delta g^{{\rm
    int},1}$, $\delta g^{{\rm int},2}$. Our results for the
interaction corrections agree with previous calculations of $\delta g^{{\rm
    int},1}$ and $\delta g^{{\rm int},2}$ by Golubev and 
Zaikin\cite{golubev2004b} for a linear array of quantum dots, and are
closely related to similar interaction corrections in a granular
metal, see Ref.\ \onlinecite{beloborodov2003b}. 
Our result for $\delta g^{\rm WL}$ agrees with
the literature in the limit of zero
temperature\cite{campagnano2006,golubev2006}
and in the high temperature limit,\cite{blanter2006} but we
are not aware of a calculation of the full temperature dependence of
$\delta g^{\rm WL}$ in the literature. (The exception is a calculation
of $\delta g^{\rm WL}$ for a double quantum dot by Golubev and Zaikin
which, however, gives an unphysical result in the limit of zero
temperature.\cite{golubev2007})

We have formulated our final results in such a way that the evaluation
of quantum corrections for a network of a relatively small number
$\ND$ of quantum dots does not require more than the inversion of an
$\ND$-dimensional matrix. All quantum corrections to the conductance
can be expressed in terms of the inter-dot conductances, form 
factors, and the
capacitances only. In principle, these parameters can be measured
independently. This makes a small quantum dot network an ideal model 
system
to compare theory and experiment, and to unambiguously identify the
mechanisms responsible for dephasing.
(Capacitances and form factors play a role only if the dots are
connected via non-ideal contacts in which one or more transmission
eigenvalues are smaller than one. For lateral quantum dot networks 
defined in semiconductor heterostructures, contacts are typically
ballistic, and the only relevant parameters are the quantized
conductances of the contacts between the quantum dots.)

The simplest example of a small quantum dot network is a `double 
quantum dot', which consists of two
quantum dots coupled to each other and to electron reservoirs via
point contacts. Several groups have reported transport measurements
on such double 
dots,\cite{waugh1995,livermore1996,oosterkamp1998,holleitner2001} 
or even on triple dots.\cite{waugh1995} (Double quantum dots also play
a prominent role in recent attempts to achieve quantum
computation.\cite{vanderwiel2003} 
However, the dots used in these experiments typically
hold one or two electrons each and can not be described by random
matrix theory.) Although, in principle,
the contact conductances in lateral double and triple
quantum dot networks are fully
tunable, the experiments of Refs.\ 
\onlinecite{waugh1995,livermore1996,oosterkamp1998,holleitner2001} 
were performed for the case that the devices
are weakly coupled to the source and drain reservoirs. In that limit,
transport is dominated by the Coulomb blockade. Our theory
applies to the opposite regime in which all dots in the network are
open, well coupled to source and/or drain reservoirs. We hope that the
availability of quantitative theoretical predictions will lead to
renewed experimental interest in quantum transport through open
quantum dots.

\section*{Acknowledgments}

We thank Igor Aleiner, Gianluigi Catelani, Jan von Delft, Simon Gravel,
Florian Marquardt, and Yuli Nazarov
for useful discussions. 
This work was supported by
the Cornell Center for Materials research under NSF grant no.\ DMR 0520404,
the Packard Foundation, the Humboldt Foundation, and by the NSF under
grant no.\ DMR 0705476. 

\appendix*

\section{Hikami Box calculation}
\label{app:hik}

In this appendix we provide details on the derivation of Eqs.\
(\ref{eq:Hikami}) and (\ref{eq:HikamiL}) of 
Sec.\ \ref{sec:model}. The explicit expression
for the Hikami box is an essential part of the calculation of the
quantum corrections to the conductance, but we have not found the
explicit expression of Eq.\ (\ref{eq:Hikami}), nor its derivation, in
the literature.

We refer to the text surrounding Eq.\ (\ref{eq:Hikami}) for the
notations used in this appendix. In general, the Hikami box
$B_{ij,kl}(\omega)$ will be nonzero only if the four indices span at
most two adjacent quantum dots. We here show the calculation of
$B_{ii,ii}(\omega)$. There are three contributions to
$B_{ii,ii}(\omega)$, which are shown in Fig.\ \ref{fg:hik}ii,ii
a--c. They read
\begin{widetext}
\begin{figure*}
  \includegraphics{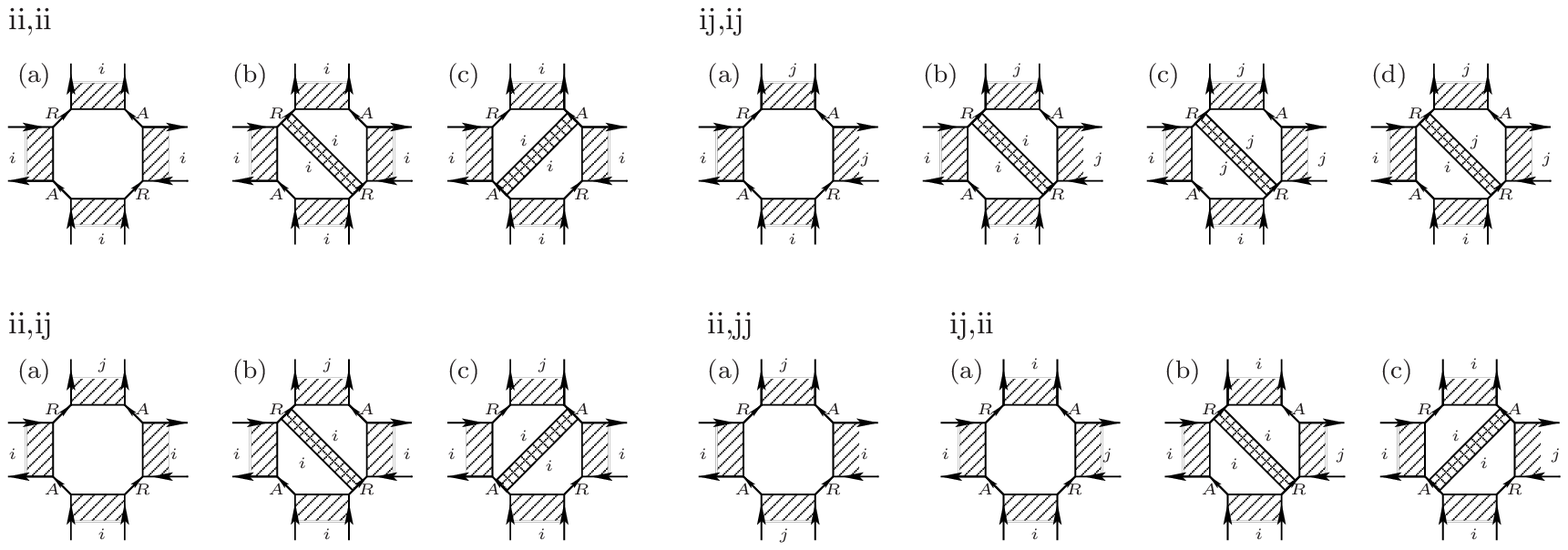}
  \caption{Diagrammatic depiction of Hikami boxes. 
  Different diagrams contribute depending on where
  the cooperon and diffuson like ladders end and begin. \label{fg:hik}}
\end{figure*}
\begin{eqnarray}
  \label{eq:Biiiia}
  B_{ii,ii}^{(a)}(\w) & =
& \tr \left[ \langle \GR_{ii}(\e_1) \rangle \langle \GA_{ii}(\e_2')
  \rangle \langle \GR_{ii}(\e_2) \rangle \langle \GA_{ii}(\e_1') \rangle 
  \right] \nonumber \\
& = & \frac{\pi^4 \nu_i^4}{\Mi^3} \left( 1 + \frac{ i  \pi \nu_i ( \e_1 - \e_1' + \e_2 - \e_2')}{2 \Mi} +  \mbox{tr}\, 
  \left[ \frac{ - 2 \Delta_i ( \Mi - \Delta_i ) }{( \Mi +
  \Delta_i)^3} + \frac{ \Delta_i^4}{\Mi ( \Mi + \Delta_i)^4}  \right]
  \right),  \\
  \label{eq:Biiiib}
  B_{ii,ii}^{(b)}(\w) &=& 
   \left( \tr \left[ \langle \GR_{ii}(\e_1) \rangle \langle \GA_{ii}(\e_2') \rangle \langle \GR_{ii}(\e_2) \right] \right)  C^{\rm RR}_{ii}(\e_1, \e_2)  \left( \tr \left[ \langle \GR_{ii}(\e_1) \rangle \langle \GR_{ii}(\e_2) \langle \GA_{ii}(\e_1') \rangle \right] \right) \nonumber \\
& = & -  \frac{\pi^4 \nu_i^4}{2 \Mi^3} \left( 1 + \frac{\tilde g_{ii} -
  \tilde g_{{\rm H},ii} + i 2 \pi ( 3\e_1 - 2 \e_1' + 3 \e_2 - 2 \e_2' )}{8 \Mi } + \frac{1}{\Mi} \tr
  \left[\frac{ \Delta_i^3 - 3 \Mi^2 \Delta_i}{ ( \Mi + \Delta_i)^3}
  \right] \right), \\
  \label{eq:Biiiic}
  B_{ii,ii}^{(c)}(\w) &=& 
\left( \tr \left[ \langle \GR_{ii}(\e_1) \rangle \langle \GA_{ii}(\e_2') \rangle \langle \GA_{ii}(\e_1') \right] \right)  C^{\rm AA}_{ii}(\e_2', \e_1')  \left( \tr \left[ \langle \GA_{ii}(\e_2') \rangle \langle \GR_{ii}(\e_2) \rangle \langle \GA_{ii}(\e_1') \right] \right) \nonumber \\
& = & -  \frac{\pi^4 \nu_i^4}{2 \Mi^3} \left( 1 + \frac{\tilde g_{ii} -
  \tilde g_{{\rm H},ii} + i 2 \pi (2 \e_1 - 3 \e_1' + 2\e_2 - 3 \e_2') }{8 \Mi } + \frac{1}{\Mi} \tr
  \left[\frac{ \Delta_i^3 - 3 \Mi^2 \Delta_i}{ ( \Mi + \Delta_i)^3}
  \right] \right),
\end{eqnarray}
\end{widetext}
where the $M_i \times M_i$ matrix $\Delta_i$ was defined in Eq.\
(\ref{eq:Deltaidef}) above and $\w = \e_1' - \e_1 + \e_2' - \e_2$. 
Traces involving the matrices $\Delta_i$
can be calculated using the identities
\begin{eqnarray}
  \label{eq:Did1}
  \tr \left[ \frac{ \Di}{(\Mi + \Di)^2} \right] 
  &  = & \sum_k \frac{ g_{ik}}{4 \Mi} ,
  \\
  \label{eq:Did2}
  \tr \left[ \frac{\Di^2}{(\Mi + \Di)^4} \right] 
  &  = & \sum_k \frac{ f_{ik}}{16 \Mi^2}.
\end{eqnarray}
Addition of Eqs.\ (\ref{eq:Biiiia})--(\ref{eq:Biiiic}) gives
\begin{eqnarray}
  B_{ii,ii}( \w ) &=& 
  \frac{ \pi^4 \nu_i^4 }{16 M_i^4} \left[ 2 \pi i \nu_i \w +
  2 ( \tilde{g}_{{\rm H},ii} + \tilde{g}_{ii} ) + \tilde{f}_{ii}
  \right].
  ~~~~ 
\end{eqnarray}

The diagrams for the relevant contributions to $B_{ij,kl}(\w)$ in
which the indices differ are shown in the other panels of Fig.\ 
\ref{fg:hik}. Expressing these contributions in terms of the matrices
$\Delta_i$ and performing the traces with the help of Eqs.\
(\ref{eq:Did1}) and (\ref{eq:Did2}), we find
\begin{eqnarray}
  B_{ij,ij}(\w) & = & 
    \frac{ \pi^4 \nu_i^2 \nu_j^2}{ 16 \Mi^2 \Mj^2 } 
    \left( f_{ij}  - g_{ij} \right), \\
  B_{ii,ij}(\w) & = & 
   \frac{ \pi^4 \nu_i^3 \nu_j }{ 16 \Mi^3 \Mj } \left( - f_{ij} \right), \\
  B_{ii,jj}(\w) & = & 
   \frac{ \pi^4 \nu_i^2 \nu_j^2 }{ 16 \Mi^2 \Mj^2 } f_{ij},
\end{eqnarray}
for $i \neq j$.
Other contributions are related by symmetry. Rewriting the general
case $B_{ij,kl}(\w)$ in terms of the matrices $\tilde g$ and $\tilde
f$ for contact conductances and form factors, we obtain the result 
given in Eq.\ (\ref{eq:Hikami}) of Sec.\ \ref{sec:model}.

\begin{figure}
  \includegraphics{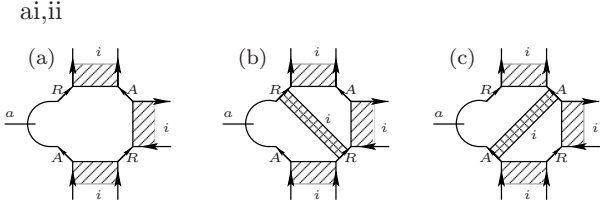}
  \caption{Diagrammatic depiction of contribution 
  from Hikami boxes placed adjacent to leads. \label{fg:hikcont}}
\end{figure}

If a Hikami box is placed adjacent to a lead, one finds the three
contributions shown in Fig.\ \ref{fg:hikcont}. Adding these we find,
with the help of Eq.\ (\ref{eq:Did2}),
\begin{eqnarray}
  B_{aj,jj}' & =& 
   \frac{\pi^5 \nu_a \nu_j^4}{ \Mj^4 }
   \tr \left[ W_{ja} W_{aj} \frac{ - \Mj^3 \Dj }{(\Mj + \Dj)^4} \right] 
 \nonumber \\ & = & 
   - \frac{ \pi^3 \nu_j^3}{16  \Mj^3}  f_{aj}'.
\end{eqnarray}
This is the result reported in Eq.\ (\ref{eq:HikamiL}) of the main text.


\begin{thebibliography}{82}
\bibitem{endnote85}
For quantum dots in semiconductor heterostructures that are defined by metal gates, 
the contact conductances are set independently by gate voltages. Each contact
 conductance can be measured by choosing gate voltages such that all other 
contacts are open, whereas the capacitances can be measured by closing off 
all contacts so that the device is in the Coulomb blockade regime. See {\protect 
\em  e.g.}, Ref.\ \protect \onlinecite{livermore1996}, 
where this procedure is used for a double quantum dot.
\end{thebibliography}
\end{document}